\documentclass[12pt]{article}

\catcode`\@=11
\@addtoreset{equation}{section}

\evensidemargin \oddsidemargin

\usepackage[dvipdfmx]{graphicx,hyperref}
\hypersetup{
setpagesize=false,
bookmarksnumbered=true,
bookmarksopen=true,
colorlinks=true,
linkcolor=black,
citecolor=black,
urlcolor=black,}

\usepackage{mathrsfs,amsbsy,amssymb,latexsym,amsfonts,amsmath,amssymb,}

 \usepackage{bm}

\usepackage[dvipdfmx]{graphicx}

\usepackage{color}

\usepackage{cite}

\newcommand\CN{\mathcal{N}}
\newcommand\CW{\mathcal{W}}

\newcommand\pa{\partial}

\newcommand\nn{\nonumber}

\newcommand\adss[2]{AdS$_{#1}\times$S$^{#2}$}

\renewcommand\d{{\rm d}}

\newcommand\bref[1]{(\ref{#1})}

\def\Fslash#1{\hspace{-0.1cm}\not\!\!#1}

\begin{document}

\vspace*{3cm}

\begin{center}
{\Large \bf 
Supersymmetric DBI Equations in Diverse Dimensions\\[0.5cm]from BRS Invariance of Pure Spinor Superstring
}
\vspace*{3cm}\\
{\large Sota Hanazawa\footnote{\texttt{s.hanazawa.aq.09@gmail.com}}
and
Makoto Sakaguchi\footnote{\texttt{makoto.sakaguchi.phys@vc.ibaraki.ac.jp}}
}
\end{center}
\vspace*{1.0cm}
\begin{center}

Department of Physics, Ibaraki University, Mito 310-8512, Japan
\end{center}

\vspace{2cm}

\begin{abstract}
We examine the BRS invariance of the open pure spinor superstring in the presence of background superfields
on a D$p$-brane. 
It is shown that the BRS invariance leads not only  to boundary conditions on the spacetime spinors,
but also to supersymmetric DBI equations of motion for the background superfields on the D$p$-brane.
These DBI equations are consistent with
the supersymmetric DBI equations for a D9-brane.
\end{abstract}

\thispagestyle{empty}
\setcounter{page}{0}

\newpage

\tableofcontents

\setcounter{footnote}{0}

\section{Introduction}
Dirac-Born-Infeld (DBI) theory is known as a non-linear generalization of Maxwell theory
and may describe, along with the Wess-Zumino action,
the low-energy effective dynamics on a single D-brane in string theory.
The bosonic DBI action is derived from the world-sheet analysis of the bosonic open string  \cite{Bosonic BI}.
A supersymmetric DBI action should be a part of
the effective action on a D-brane in type II superstring theory. 
In the Ramond-Neveu-Schwarz (RNS) formulation,
however,
it is difficult 
to read off the target space geometry coupling to Ramond-Ramond fields,
because space-time supersymmetry becomes manifest only after the GSO projection.
So the RNS superstring has led to the only bosonic sector of the supersymmetric DBI action \cite{RNS BI}.

\medskip

The Green-Schwarz (GS) formulation has an advantage in this direction.
The Wess-Zumino term
which ensures the $\kappa$-invariance of the world-volume action of a D-brane
 is constructed in \cite{brane kappa WZ}.
In \cite{Cayley image}, the $\kappa$-symmetric approach, so called the superembedding formalism \cite{Origin embed},
is shown to lead to linearised supersymmetric DBI equations of motion for a D9-brane,
which have the ten-dimensional ${\mathcal N}=2$ supersymmetry.
Furthermore, in \cite{GS open background}, 
the classical $\kappa$-invariance of
 an open GS superstring in an abelian background
 is shown to imply that the background fields should satisfy full non-linear equations of motion for a supersymmetric DBI action.
Non-abelian extension of this formalism is
discussed in \cite{boundary fermion}
as the boundary fermion formalism 
where Chan-Patton factors describing coincident D-branes are replaced by boundary fermions\footnote
{Other than this study, there have been many attempts to extend to the non-abelian DBI theory based on 
various approaches 
\cite{NADBI BRS  87,NADBI T 97,NADBI TR 00,NADBI K 00,NADBI STT 01,NADBI BRS 01,NADBI RSTZ 01,NADBI S 01,NADBI KS 01,NADBI CRE 02,NADBI MBM 02,NADBI DHL 02,NADBI DHHK 03,NADBI CM 03,NADBI BM 05}.}. 

\medskip

Unlike the formulations mentioned above,
the pure spinor formulation \cite{PS} enables us to
quantize a superstring in a super-Poincar\'e covariant manner.
In this formulation,
the $\kappa$-symmetry in the GS formulation is replaced with the BRS symmetry.
It is shown 
in  \cite{PS open background},
correspondingly to the $\kappa$-symmetry analysis \cite{GS open background}, 
that
the classical BRS invariance of an open pure spinor superstring  leads to
 supersymmetric DBI equations of motion on a D9-brane,
which have the non-linear ${\cal N}=1$ supersymmetry
as well as
the manifest ${\cal N}=1$ supersymmetry.
These equations precisely coincide with those obtained in the superembedding formalism \cite{Ker D9 DBI}.
Furthermore,
the non-abelian extension of supersymmetric DBI equations is proposed.
In \cite{Wyl D brane} (see also \cite{Gra D-brane} and \cite{Muk D brane}),
D-brane boundary states are constructed in the pure spinor formulation.
Especially, calculating the disk scattering amplitude suggests
that the coupling of the boundary state to the background fields will reproduce the DBI kinetic term and the Wess-Zumino term of the D9-brane effective action.
These achievements might imply the fact 
that the low-energy effective theory on the D9-brane is determined uniquely by the ten-dimensional ${\mathcal N}=2$ supersymmetry.

\medskip

In this paper we will derive supersymmetric DBI equations of motion on a D$p$-brane,
as well as
a D9-brane,
from the BRS invariance
of the open pure spinor superstring.
Our approach is similar to that
taken 
in \cite{PS open background} for a D9-brane.
However
the inclusion of Dirichlet components requires
improvements which are not just a dimensional reduction
of the case of a D9-brane.
As in \cite{PS open background}, we will provide two boundary terms,
the counter term $S_{\rm b}$ for the ${\mathcal N}=1$ supersymmetry transformation of the world-sheet action $S_0$ 
and the background superfield coupling $V$ as a relevant extension of the pure spinor vertex operator.
It is found that
the contribution of Dirichlet components in them cannot be determined unless considering the BRS invariance.
In \cite{PS open background},
by using non-trivial boundary conditions
given by the general variation of $S_0+S_{\rm b}+V$, 
the BRS charge conservation leads to supersymmetric DBI equations for the D9-brane.
On the other hand, we will show that supersymmetric DBI equations in diverse dimensions
are extracted  only from the BRS transformation of $S_0+S_{\rm b}+V$ under an identification on the D-brane position.

\medskip

This paper is organized as follows. In section \ref{section 2}, after introducing the type II pure spinor open superstring action, we construct the boundary term for the ${\mathcal N}=1$ supersymmetry invariance of this action.
In section \ref{Background action}, 
background superfield coupling  is found by considering the modification of a vertex operator
in the open pure spinor superstring.
In section \ref{section BRS}, 
we confirm that these background superfields satisfy supersymmetric DBI equations of motion. 
The last section is devoted to summary and discussions. 
In addition, we give a brief review of the covariant approach 
for the ten-dimensional ${\cal N}= 1$ super Yang-Mills theory
in Appendix \ref{appendix:SYM}.
We will formulate a vertex operator in the open pure spinor superstring in Appendix \ref{appendix:vertex}.
We show that our result  can be derived  also from improving the method used in \cite{PS open background} 
to include Dirichlet components in Appendix \ref{appendix:charge conservation}.

\section{Open pure spinor superstring} \label{section 2}

The world-sheet action of the type II pure spinor open superstring \cite{PS} is
given as
\begin{align}
S_0=\frac{1}{\pi\alpha'}\int {\d z}{\d \bar{z}}
\left\{\frac{1}{2}{\partial}x^m\bar{\partial}x_m
+p_{\alpha}\bar{\partial}\theta^{\alpha}+\widehat{p}_{\alpha}\partial\widehat{\theta}^{\alpha}
+\omega_{\alpha}\bar{\partial}\lambda^{\alpha}+\widehat{\omega}_{\alpha}\partial\widehat{\lambda}^{\alpha} \right\}~,
\label{PS action}
\end{align}
where $x^m$ ($m=0,1,\cdots,9$) is a ten-dimensional coordinate,
$\theta^{\alpha}$ and $\widehat{\theta}^{\alpha}~(\alpha=1,\cdots,16)$ are left- and right-moving ten-dimensional Majorana-Weyl spinors,
respectively,
and 
$\lambda^{\alpha}$ and $\widehat{\lambda}^{\alpha}$ are bosonic ghosts satisfying pure spinor constrains 
$\lambda\gamma^m\lambda=\widehat{\lambda}\gamma^m\widehat{\lambda}=0$\,.
The $(p_\alpha,\widehat{p}_\alpha)$ and $(\omega_\alpha,\widehat{\omega}_\alpha)$
are conjugate to $(\theta^\alpha,\widehat{\theta}^\alpha)$
 and $(\lambda^\alpha,\widehat{\lambda}^\alpha)$, respectively.
The world-sheet derivatives
$\partial$ and $\bar{\partial}$ denote
$\partial=\partial_{\tau}+\partial_{\sigma}$ 
and 
$\bar{\partial}=\partial_{\tau}-\partial_{\sigma}$, 
respectively.
It  implies ${\d z}{\d \bar{z}}=-\frac{1}{2}{\d \tau}{\d \sigma}$. 
The action is invariant under the 
gauge transformations 
$\delta_{\Lambda}\omega_{\alpha}=\Lambda^m(\gamma_m\lambda)_{\alpha}$ and 
$\delta_{\widehat{\Lambda}}\widehat{\omega}_{\alpha}=\widehat{\Lambda}^m(\gamma_m\widehat{\lambda})_{\alpha}$. 
We use
$16\times16$ symmetric matrices $\gamma^m_{\alpha\beta}$ and $\gamma^{m\alpha\beta}$ 
which are off-diagonal blocks of the $32\times32$ gamma matrices 
and satisfy $\gamma^m_{\alpha\beta}\gamma^{n\beta\gamma}+\gamma^n_{\alpha\beta}\gamma^{m\beta\gamma}=2\eta^{mn}\delta_{\alpha}^{\gamma}$.
We frequently use the Fierz identity $\gamma_{m(\alpha\beta}\gamma^m_{\gamma)\delta}=0$. 

The action (\ref{PS action}) is invariant under the ten-dimensional ${\mathcal N}=2$ supersymmetry transformations
\begin{eqnarray}
\begin{split}
&\delta_{\epsilon}\theta^{\alpha}=\epsilon^{\alpha}~,~~
\delta_{\epsilon}\widehat{\theta}^{\alpha}=\widehat{\epsilon}^{\alpha}~,~~
\delta_{\epsilon}x^m=\frac{1}{2} \theta\gamma^m\epsilon
+\frac{1}{2} \widehat{\theta}\gamma^m\widehat{\epsilon}~,\\[0.2cm]
&\delta_{\epsilon}p_{\alpha}=\frac{1}{2}{\partial}x^m(\gamma_m\epsilon)_{\alpha}
-\frac{1}{8}(\epsilon\gamma^m\theta)(\gamma_m\partial\theta)_{\alpha}~,~~
\delta_{\epsilon}\widehat{p}_{\alpha}=\frac{1}{2}\bar{\partial}x^m(\gamma_m\widehat{\epsilon})_{\alpha}
-\frac{1}{8}(\widehat{\epsilon}\gamma^m\widehat{\theta})(\gamma_m\bar{\partial}\widehat{\theta})_{\alpha}~,
\end{split}
\label{N2 SUSY}
\end{eqnarray}
where parameters $\epsilon$ and $\widehat\epsilon$ 
correspond to ten-dimensional Majorana-Weyl spinors. 
For an open superstring, we are left with a surface term
\begin{align}
\delta_{\epsilon}S_0=\frac{1}{2\pi\alpha'}\int\d\tau\left\{
\frac{1}{2}(\epsilon\gamma^m\theta-\widehat{\epsilon}\gamma^m\widehat{\theta})\dot{x}_m
+\frac{1}{12}(\epsilon\gamma^m\theta)(\theta\gamma_m\dot{\theta})
-\frac{1}{12}(\widehat{\epsilon}\gamma^m\widehat{\theta})
(\widehat{\theta}\gamma_m\dot{\widehat{\theta}})
\right \}\Biggl|
~, 
\label{SUSY surface}
\end{align}
where ``$|
$'' means ``evaluated at the boundary'' and we will omit it for brevity
in the following.
A dot on a field denotes the $\tau$-derivative of the field,
while a prime does the $\sigma$-derivative.
If there are no background fields, the surface term (\ref{SUSY surface}) can be eliminated by imposing usual boundary conditions for D$p$-branes\footnote
{We must impose the same boundary condition on $\theta$ and $\lambda$ since BRS transformations relate them each other. These boundary conditions also eliminate the surface term which comes from the BRS transformation of the world-sheet action $S_0$\,. 
See \cite{HS AdS, HS mem} for related topics.}. 
\begin{align}
x'^{\mu}=0~,~~\dot{x}^i=0~,~~
\widehat{\theta}=\gamma^{1 \cdots p}\theta~,~~
\widehat{\lambda}=\gamma^{1 \cdots p}\lambda~,
\end{align}
and $\CN=1$ supersymmetry condition $\widehat{\epsilon}=\gamma^{1 \cdots p}\epsilon$.
These boundary conditions imply that $p=$ odd for  the type IIB string while $p=$ even for the  type IIA string.
As usual, $x^{\mu}~(\mu=0,\cdots,p)$ are Neumann coordinates,
while  $x^{i}~(i=p+1,\cdots,9)$ are Dirichlet coordinates.

Instead of imposing boundary conditions,
we will consider coupling to the background superfields
preserving the $\CN=1$ supersymmetry specified by $\widehat{\epsilon}=\gamma^{1 \cdots p}\epsilon$.
To preserve  $\CN=1$ supersymmetry,
we must  introduce a boundary term which eliminates  (\ref{SUSY surface}).


\section{$\CN=1$ supersymmetry and boundary term}

Here we will introduce a boundary term $S_{\rm b}$ 
which leaves $S_0+S_{\rm b}$ invariant under the $\CN=1$ 
supersymmetry.
For this purpose, it is convenient to introduce the following objects
\begin{eqnarray}
\begin{split}
&\theta_{\pm}^{\alpha}=\frac{1}{\sqrt{2}}\left(\widehat{\theta}^{\alpha}\pm(\gamma^{1{\cdots}p}\theta)^{\alpha} \right)~,~~
&&d^{\pm}_{\alpha}=\sqrt{2}\left(\widehat{d}_{\alpha}\pm(\gamma^{1{\cdots}p}d)_{\alpha} \right)~,\\
&\lambda^{\alpha}_{\pm}=\frac{1}{\sqrt{2}}\left(\widehat{\lambda}^{\alpha}\pm(\gamma^{1{\cdots}p}\lambda)^{\alpha} \right)~,~~
&&\omega_{\alpha}^{\pm}=\sqrt{2}\left(\widehat{\omega}_{\alpha}\pm(\gamma^{1{\cdots}p}\omega)_{\alpha} \right)~,
\end{split}
\end{eqnarray}
where 
$d_{\alpha}=p_{\alpha}-\frac{1}{2}{\partial}x^m(\gamma_m\theta)_{\alpha}-\frac{1}{8}(\theta\gamma^m\partial\theta)(\gamma_m\theta)_{\alpha}$ and $\widehat{d}_{\alpha}=\widehat{p}_{\alpha}-\frac{1}{2}\bar{\partial}x^m(\gamma_m\widehat{\theta})_{\alpha}-\frac{1}{8}(\widehat{\theta}\gamma^m\bar{\partial}\widehat{\theta})(\gamma_m\widehat{\theta})_{\alpha}$
 are invariant under 
the $\epsilon$- and $\widehat\epsilon$-supersymmetry in \bref{N2 SUSY},
respectively.
By using these variables, the ${\mathcal N}=1$ supersymmetry transformations
specified by $\widehat{\epsilon}=\gamma^{1 \cdots p}\epsilon$
are represented as
\begin{align}
&\delta_{\eta}\theta_+^{\alpha}=\eta^{\alpha}~,~~~
\delta_{\eta}\theta_-^{\alpha}=0~,~~~
\delta_{\eta}{x}^{\mu}=\frac{1}{2}\,{\theta}_{+}\gamma^{\mu}\eta~,~~~
\delta_{\eta}{x}^{i }=\frac{1}{2}\,{\theta}_{-}\gamma^{i }\eta~,~~~
\delta_{\eta}\lambda_{\pm}^{\alpha}
=\delta_{\eta}\omega^{\pm}_{\alpha}=0~,
\label{N1 SUSY}
\end{align}
where we introduced 
$\eta$ by $\eta\equiv \frac{1}{\sqrt{2}}(\widehat{\epsilon}+\gamma^{1 \cdots p}\epsilon)$.
The $\CN=1$ supersymmetry transformation of $S_0$
is found to be
\begin{align}
\delta_{\eta}S_0=-\frac{1}{2\pi\alpha'}\int {\d \tau}&\left\{
\frac{1}{2}(\eta\gamma^{\mu}\theta_-)\dot{x}_{\mu}
+\frac{1}{2}(\eta\gamma^{i }\theta_+)\dot{x}_{i }
 \right.
\nn\\
&\left.
+\,\frac{1}{8}(\eta\gamma^m\theta_+)(\theta_-\gamma_m\dot{\theta}_+)
+\frac{1}{24}(\eta\gamma^m\theta_-)(\theta_-\gamma_m\dot{\theta}_-)
 \right\}~,
\label{N1 SUSY surface term}
\end{align}
where we have used  the Fierz identity.

The boundary term $S_{\rm b}$
we found
is
\begin{align}
S_{\mathrm b}
=\frac{1}{2\pi\alpha'}\int {\rm d} \tau & \left\{
\frac{1}{2}\Pi_+^{\mu}(\theta_+\gamma_{\mu}\theta_-)
-\frac{1}{2}y^i(\theta_+\gamma_{i } \dot{\theta}_+)
-\frac{1}{8}(\theta_+\gamma^{\mu}\theta_-)(\theta_+\gamma_{\mu}\dot{\theta}_+)
+\frac{1}{8}(\theta_+\gamma^i\theta_-)(\theta_+\gamma_i\dot{\theta}_+) \right.
\nn\\
&\left.
+\,\frac{1}{24}(\theta_+\gamma^m\theta_-)(\theta_-\gamma_m\dot{\theta}_-)
+\frac{1}{2}c_1\Delta_{\alpha}^+\theta^{\alpha}_-+\frac{1}{2}c_2\omega^+_{\alpha}\lambda_-^{\alpha}
+y_i\widetilde{\Pi}_+^i
 \right\}~,
\label{SUSY counter term}
\end{align}
where $c_1$ and $c_2$ are constants.
We have introduced the followings
\begin{eqnarray}
\begin{split}
&
\begin{split}
&\Pi_{+}^{\mu}
=\frac{1}{2}\left(\widehat{\Pi}^{\mu}+\Pi^{\mu} \right)
-\frac{1}{2}(\theta_-\gamma^{\mu}\dot{\theta}_- )
~,~~~~
&&\Pi_{-}^i
=\frac{1}{2}\left(\widehat{\Pi}^i+\Pi^i \right)
-\frac{1}{2}(\theta_-\gamma^i\dot{\theta}_+ )~,
\\
&\widetilde{\Pi}^{\mu}_{-}
=\frac{1}{2}\left(\widehat{\Pi}^{\mu}-\Pi^{\mu} \right)
-\frac{1}{2}(\theta_-\gamma^{\mu}\dot{\theta}_+ )~,~~~
&&\widetilde{\Pi}^i_{+}
=\frac{1}{2}\left(\widehat{\Pi}^i-\Pi^i \right)
-\frac{1}{2}(\theta_-\gamma^i\dot{\theta}_- )~,
\end{split}
\\
&y^i=x^i+\frac{1}{2}(\theta_+\gamma^i\theta_- )~,
\\
&
\Delta^+_{\alpha}=d_{\alpha}^{+}
+\frac{1}{2}(\gamma^{\mu}\theta_{-})_{\alpha}\left(\widehat{\Pi}_{\mu}-\Pi_{\mu} \right)
+\frac{1}{2}(\gamma^{i }\theta_{-})_{\alpha}\left(\widehat{\Pi}_{i }+\Pi_{i } \right)~,
\end{split}
\label{def y}
\end{eqnarray}
where
$\Pi^m={\partial}x^m+\frac{1}{2}\theta\gamma^m\partial\theta$ 
and 
$\widehat{\Pi}^m=\bar{\partial}x^m+\frac{1}{2}\widehat{\theta}\gamma^m\bar{\partial}\widehat{\theta}$ 
are $\epsilon$- and $\widehat\epsilon$-supersymmetry invariants, respectively.
Objects in \bref{def y}
are invariant under the $\CN=1$ supersymmetry.
To show this,
we have to treat objects like $\theta_\pm'$ at the boundary.
For this, we require that
at the boundary
\begin{align}
\theta'_\pm=-\dot\theta_\mp~,~~~
\lambda'_\pm=-\dot\lambda_\mp~.
\end{align}
These are consistent with the bulk equations of motion
$\bar{\partial}\theta^{\alpha}=\partial\widehat{\theta}^{\alpha}=\bar{\partial}\lambda^{\alpha}=\partial\widehat{\lambda}^{\alpha}=0$.
It is shown that this choice leads to DBI equations in this paper.
We also note that
the last three terms in (\ref{SUSY counter term})
are invariant under the $\CN=1$ supersymmetry separately.
This implies that  they are not determined from the $\CN=1$ supersymmetry.
It is worth noting that
\bref{SUSY counter term} cannot be extracted as a
dimensional reduction of the one for the D9-brane.

\subsection{BRS symmetry} \label{sec:BRS}

We shall show that the last term $y_i\widetilde{\Pi}_+^i$ in \bref{SUSY counter term}
is required by the BRS invariance of $S_0+S_\mathrm{b}$,
when
there is no background superfield coupling.

The action \bref{PS action} is invariant under a pair of BRS variations,
say $\delta_1$ and $\delta_2$.
In the presence of the boundary,
these BRS variation must satisfy $\delta_1=\delta_2$ at the boundary.
This implies that the BRS transformations $\delta_Q=\delta_1+\delta_2$
remain unbroken in the presence of the boundary
\begin{align}
&\delta_Q\theta_{\pm}^{\alpha}=\lambda_{\pm}^{\alpha}~,~~~~\delta_Q\lambda_{\pm}^{\alpha}=0~,~~~~\delta_Q\omega^{\pm}_{\alpha}=d^{\pm}_{\alpha}~,
\nn\\
&\delta_Qx^{\mu}=\frac{1}{2}\lambda_+\gamma^{\mu}\theta_++\frac{1}{2}\lambda_-\gamma^{\mu}\theta_-~,~~~~
\delta_Qx^{i}=\frac{1}{2}\lambda_+\gamma^{i}\theta_-+\frac{1}{2}\lambda_-\gamma^{i}\theta_+~,~~~~
\delta_Qy^i=\lambda_+\gamma^i\theta_-~,
\nn\\
&\delta_Q\Pi_+^{\mu}=
\lambda_+\gamma^{\mu}\dot{\theta}_++\frac{1}{2}\lambda_-\gamma^{\mu}\dot{\theta}_-+\frac{1}{2}\dot{\lambda}_-\gamma^{\mu}\theta_-~,~~~~
\delta_Q\Pi_-^{i}=
\lambda_+\gamma^{i}\dot{\theta}_-+\frac{1}{2}\lambda_-\gamma^{i}\dot{\theta}_++\frac{1}{2}\dot{\lambda}_+\gamma^{i}\theta_-~,
\nn\\
&\delta_Q\widetilde{\Pi}^{\mu}_{-}=
\lambda_+\gamma^{\mu}\dot{\theta}_-+\frac{1}{2}\lambda_-\gamma^{\mu}\dot{\theta}_++\frac{1}{2}\dot{\lambda}_+\gamma^{\mu}\theta_-~,~~~~
\delta_Q\widetilde{\Pi}^i_{+}=
\lambda_+\gamma^{i}\dot{\theta}_++\frac{1}{2}\lambda_-\gamma^{i}\dot{\theta}_-+\frac{1}{2}\dot{\lambda}_-\gamma^{i}\theta_-~,
\nn\\
&\delta_Q\Delta_{\alpha}^+=
-2(\gamma_{\mu}\lambda_+)_{\alpha}\Pi_+^{\mu}-2(\gamma_{i}\lambda_+)_{\alpha}\widetilde{\Pi}_+^{i}
-(\gamma_{\mu}\lambda_-)_{\alpha}\widetilde{\Pi}_-^{\mu}-(\gamma_{i}\lambda_-)_{\alpha}\Pi_-^{i}
\nn\\
&~~~~~~~~~~~~-(\gamma^m\dot{\theta}_-)_{\alpha}(\lambda_+\gamma_m\theta_-)-(\gamma^m\theta_-)_{\alpha}(\lambda_-\gamma_m\dot{\theta}_+)
-\frac{1}{2}(\gamma^m\lambda_-)_{\alpha}(\theta_-\gamma_m\dot{\theta}_+)~.
\label{BRS transformationsII}
\end{align}
Again, we find the world-sheet action $S_0$ is BRS invariant $\delta_QS_0=0$ up to a surface term,
and satisfies
\begin{align}
{\delta}_Q(S_0+S_{\rm b})=\frac{1}{2\pi\alpha'}\int {\rm d}\tau & \biggl\{
(1-c_1)\Pi_+^{\mu}(\lambda_+\gamma_{\mu}\theta_-)
-\frac{1}{2}(c_1+c_2)\Pi_-^i(\lambda_-\gamma_i\theta_-)
\nn\\
&-\frac{1}{2}(c_1+c_2)\widetilde{\Pi}_-^{\mu}(\lambda_-\gamma_{\mu}\theta_-) 
+(1-c_1)\widetilde{\Pi}_+^i(\lambda_+\gamma_i\theta_-)
\nn\\
&+\frac{1}{2}(c_2-c_1)\Delta^+_{\beta}\lambda_-^{\beta}
+\Bigl(-\frac{1}{3}+\frac{c_1+c_2}{4} \Bigr)
(\lambda_-\gamma^{m}\theta_-)
(\dot{\theta}_+\gamma_{m}\theta_-)
\nn\\
&+\frac{1}{2}\Bigl(\frac{1}{3}-c_1 \Bigr)
(\lambda_+\gamma^m\theta_-)
(\theta_-\gamma_m\dot{\theta}_-)
\biggr\}~.
\label{BRS surface}
\end{align}
Let us assume that there are no background fields.
In this case,
the 
\bref{BRS surface} must be eliminated by the usual boundary conditions
$\theta_-^{\alpha}=\lambda_-^{\alpha}=0$.
It is obvious to see that these boundary conditions eliminate \bref{BRS surface}  as expected.
It should be noted that this happens only when
we include the term $y_i\widetilde{\Pi}_+^i$
in \bref{SUSY counter term}.

Finally we comment on $y_i$.
Remarkably, the BRS transformation of $S_0+S_{\rm b}$, at the boundary, is independent of $y_i$. 
More generally, we confirm ${\delta}(S_0+S_{\rm b})/{\delta y_i}\Bigr|=0$ in Appendix \ref{appendix:charge conservation}.
This strongly suggests that $y_i$ should represent the position of the D$p$-brane.


\section{Supersymmetric DBI equations of motion}\label{section 4}

In this section, we will give the background coupling $V$ in terms of superfields
on a D$p$-brane.
Examining the BRS variation of $S_0+S_{\rm b}+V$, we obtain supersymmetric DBI equations of motion
on the D$p$-brane.

\subsection{Background superfield coupling for D$p$-branes \label{Background action}}

In Appendix A,
 we define the ten-dimensional $\CN=1$ superfield $A_M=(A_m,A_\alpha)$.
We introduce background superfields
on a D$p$-brane as
a dimensional reduction of $A_M$:
$A_m=(A_\mu(x^\mu,\theta_+),A_i(x^\mu,\theta_+))$
and
$A_\alpha=A_\alpha(x^\mu,\theta_+)$.
Obviously they are invariant under the $\CN=1$ supersymmetry.
Similarly we introduce $\CW^\alpha=\CW^\alpha(x^\mu,\theta_+)$ and
$F_{mn}=F_{mn}(x^\mu,\theta_+)$\,.
We use the ten-dimensional Majorana-Weyl spinor notation throughout this paper.
This means that
we are considering the DBI equations with 16 supersymmetries,
for example $\CN=4$ supersymmetric DBI equations on a D3-brane.
 
The background coupling $V$
used in \cite{PS open background} is regarded as an extension of the vertex operator of
the open pure spinor superstring.
We give a brief review of the vertex operator in Appendix \ref{appendix:vertex}.

The background coupling $V$ we introduce is
\begin{align}
V=\frac{1}{2\pi\alpha'}\int{\rm d}\tau & \Bigl\{\dot{\theta}_{+}^{\alpha}A_{\alpha}(x^{\mu},\theta_{+})
+\Pi_{+}^{\mu}A_{\mu}(x^{\mu},\theta_{+})
+\widetilde{\Pi}_{+}^{i }A_{i }(x^{\mu},\theta_{+})
\Bigr.
\nonumber 
\\
&~~~~~~~~~~~~~~~~
+\frac{1}{2}\Delta^{+}_{\alpha}{\cal W}^{\alpha}(x^{\mu},\theta_{+})
+\frac{1}{4}N_{+}\Fslash{\mathcal F}(x^{\mu},\theta_{+})
\Bigr\}
\label{background action}
\end{align}
where
\begin{align}
(N_+)^{\,\beta}_{\alpha}
=\frac{1}{2}\omega^+_{\alpha}\lambda_+^{\beta}
~,~~~
\Fslash{\mathcal F}^{\alpha}_{~\beta}
=\delta^{\alpha}_{\,\beta}{\cal F}^{(0)}
+(\gamma^{mn})^{\alpha}{}_{\beta}{\cal F}^{(2)}_{mn}
+(\gamma^{mnpq})^{\alpha}{}_{\beta}{\cal F}^{(4)}_{mnpq}~.
\end{align}
Note that ${\cal F}^{(0)}$, ${\cal F}_{mn}^{(2)}$ and ${\cal F}_{mnpq}^{(4)}$ are some possible products of any number of vector field strengths ${F}_{mn}$ \footnote
{There are no more higher forms because of the property 
\begin{align}
\omega_\alpha(\gamma^{m_1 \cdots m_{2k}})^{\alpha}{}_{\beta}\lambda^\beta
=\pm\frac{1}{(10-2k)!}(-1)^{k+1}\epsilon^{m_1 \cdots m_{2k}}{}_{\,n_1 \cdots n_{10-2k}}\omega_\alpha(\gamma^{n_1 \cdots n_{10-2k}})^{\alpha}{}_{\beta}\lambda^\beta~,
\end{align}
where the sign in the right hand side
depends on the chirality of $\lambda$.}, 
 which is consistent with analysis for D-brane boundary states \cite{Wyl D brane} from the viewpoint of the pure spinor closed superstring. 
Needless to say, the $V$ is invariant under the $\CN=1$ supersymmetry.
Since we have made the factor $1/(2\pi\alpha')$ manifest in $V$, 
dimensions of these superfields differ from conventional ones.
In this sense, we assign dimensions to $[A_{\alpha}]$, $[A_m]$, $[{\cal W}^{\alpha}]$ and 
$[{F}_{mn}]$ as $-\frac{3}{2}$, $-1$, $-\frac{1}{2}$ and $0$, respectively.

\subsection{DBI equations from BRS symmetry}
\label{section BRS}

In this subsection, 
we will add the background superfield coupling $V$ in \bref{Background action} to the action $S_0+S_{\rm b}$ 
and then require that the BRS variation
 $\delta_Q(S_0+S_{\rm b}+V)$ vanishes. 
 This requirement leads to boundary conditions on spacetime spinors and conditions on background superfields.
The latter is found to be supersymmetric DBI equations of motion for them.
 
We find that the BRS variation $\delta_Q V$ may be expressed as
\begin{align}
{\delta}_QV=\frac{1}{2\pi\alpha'}\int & {\rm d}\tau\biggl\{
\Pi_+^{\mu}\Bigl[-\lambda_+^{\alpha}\partial_{\mu}A_{\alpha}+\lambda_+^{\alpha}D_{\alpha}A_{\mu}
+\frac{1}{2}(\lambda_-\gamma^{n}\theta_-)(\partial_{n}A_{\mu}-\partial_{\mu}A_{n})
-(\lambda_+\gamma_{\mu}{\cal W})\, \Bigr]
\nn\\
&+\Pi_-^i\Bigl[-\frac{1}{2}(\lambda_-\gamma_i{\cal W})
-\frac{1}{8}(\gamma_i\theta_-)_{\alpha}\lambda_+^{\beta}\Fslash{\mathcal F}^{\alpha}_{~\beta} \Bigr]
\nn\\
&+\widetilde{\Pi}_-^{\mu}\Bigl[-\frac{1}{2}(\lambda_-\gamma_{\mu}{\cal W})
-\frac{1}{8}(\gamma_{\mu}\theta_-)_{\alpha}\lambda_+^{\beta}\Fslash{\mathcal F}^{\alpha}_{~\beta} \Bigr]
\nn\\
&+\widetilde{\Pi}_+^{i}\Bigl[-(\lambda_+\gamma_i{\cal W})+\lambda_+^{\alpha}D_{\alpha}A_i+\frac{1}{2}(\lambda_-\gamma^{\mu}\theta_-)\partial_{\mu}A_i \Bigr]
\nn\\
&+\frac{1}{2}\Delta^+_{\beta}\Bigl[-\lambda_+^{\alpha}D_{\alpha}{\cal W}^{\beta}-\frac{1}{2}(\lambda_-\gamma^{\mu}\theta_-)\partial_{\mu}{\cal W}^{\beta}
+\frac{1}{4}\lambda_+^{\alpha}\Fslash{\mathcal F}^{\beta}_{~\alpha} \Bigr]
\nn\\
&+\frac{1}{4}{N_{+\beta}}^{\gamma}\Bigl[\lambda_+^{\alpha}D_{\alpha}\Fslash{\mathcal F}^{\beta}_{~\gamma}
+\frac{1}{2}(\lambda_-\gamma^{\mu}\theta_-)\partial_{\mu}\Fslash{\mathcal F}^{\beta}_{~\gamma} \Bigr]
\nn\\
&+\dot{\theta}_+^{\beta}\Bigl[-\lambda_+^{\alpha}D_{\beta}A_{\alpha}-\lambda_+^{\alpha}D_{\alpha}A_{\beta}
-\frac{1}{2}(\lambda_-\gamma^{\mu}\theta_-)\partial_{\mu}A_{\beta}+(\gamma^{m}\lambda_+)_{\beta}A_{m}
\nn\\
&~~~~+\frac{1}{2}(\lambda_-\gamma^{m}\theta_-)D_{\beta}A_{m}
+\frac{1}{2}(\gamma^m\lambda_-)_{\beta}(\theta_-\gamma_m{\cal W})
\nn\\
&~~~~+\frac{1}{4}(\gamma^m\theta_-)_{\beta}(\lambda_-\gamma_m{\cal W})
+\frac{1}{16}(\gamma^m\theta_-)_{\beta}(\gamma_m\theta_-)_{\gamma}\lambda_+^{\alpha}\Fslash{\mathcal F}^{\gamma}_{~\alpha} \,\Bigr]
\nn\\
&+\dot{\theta}_-^{\beta}\Bigl[-\frac{1}{2}(\lambda_+\gamma^m\theta_-)(\gamma_m{\cal W})_{\beta} \Bigr]~ \biggr\}~.
\label{deltaQ v}
\end{align}
Note that the supercovariant derivative on the D$p$-brane is defined by
\begin{align}
D_{\alpha}=\frac{\partial}{\partial\theta_+^{\alpha}}
+\frac{1}{2}(\gamma^{\mu}\theta_+)_{\alpha}\partial_{\mu}~.
\label{supercovariant derivative}
\end{align}
Gathering (\ref{BRS surface}) and (\ref{deltaQ v}) together, we obtain the BRS variation of 
{$S_0+S_{\rm b}+V$}
 as
\begin{align}
{\delta}_Q(S_0+S_{\rm b}+V)=\frac{1}{2\pi\alpha'}\int {\rm d}\tau & \biggl\{
\Pi_+^{\mu}{\rm X}_{\mu}
+\widetilde{\Pi}_+^i{\rm X}_i 
+\Pi_-^i{\rm Y}_i
+\widetilde{\Pi}_-^{\mu}{\rm Y}_{\mu} 
\nn \\
&-\frac{1}{2}\Delta^+_{\beta}\Lambda^{\beta}
+\frac{1}{4}{N_{+\beta}}^{\alpha}{\rm Z}_{~\alpha}^{\beta}
+\dot{\theta}_+^{\alpha}\Theta^+_{\alpha}+\dot{\theta}_-^{\alpha}\Theta^-_{\alpha} \biggr\}~,
\label{Total BRS surface}
\end{align}
where ${\rm X}_m$, ${\rm Y}_m$, $\Lambda^{\beta}$, ${\rm Z}_{~\alpha}^{\beta}$ and $\Theta^{\pm}_{\alpha}$ are given as follows
\begin{align}
&{\rm X}_m \equiv (1-c_1)(\lambda_+\gamma_m\theta_-)-\lambda_+^{\alpha}\partial_mA_{\alpha}+\lambda_+^{\alpha}D_{\alpha}A_m
-\frac{1}{2}(\lambda_-\gamma^{n}\theta_-)(\partial_mA_n-\partial_nA_m)
\nn\\
&~~~~~~
-(\lambda_+\gamma_m{\cal W})~,
\\
&{\rm Y}_m \equiv -\frac{1}{2}(c_1+c_2)(\lambda_-\gamma_m\theta_-)-\frac{1}{2}(\lambda_-\gamma_m{\cal W})
-\frac{1}{8}(\gamma_m\theta_-)_{\alpha}\lambda_+^{\beta}\Fslash{\mathcal F}_{~\beta}^{\alpha}~,
\\
&\Lambda^{\beta} \equiv (c_1-c_2)\lambda_-^{\beta}+\lambda_+^{\alpha}D_{\alpha}{\cal W}^{\beta}+\frac{1}{2}(\lambda_-\gamma^{\mu}\theta_-)\partial_{\mu}{\cal W}^{\beta}
-\frac{1}{4}\lambda_+^{\alpha}\Fslash{\mathcal F}_{~\alpha}^{\beta}~,
\\
&{\rm Z}_{~\alpha}^{\beta} \equiv \lambda_+^{\gamma}D_{\gamma}\Fslash{\mathcal F}_{~\alpha}^{\beta}
+\frac{1}{2}(\lambda_-\gamma^{\mu}\theta_-)\partial_{\mu}\Fslash{\mathcal F}_{~\alpha}^{\beta}~,
\\
&\Theta^+_{\alpha} \equiv \Bigl(-\frac{1}{3}+\frac{c_1+c_2}{4} \Bigr)
(\gamma^{m}\theta_-)_{\alpha}(\lambda_-\gamma_{m}\theta_-)
-\lambda_+^{\beta}(D_{\alpha}A_{\beta}+D_{\beta}A_{\alpha})
-\frac{1}{2}(\lambda_-\gamma^{\mu}\theta_-)\partial_{\mu}A_{\alpha}
\nn\\
&~~~~~~
+(\gamma^{m}\lambda_+)_{\alpha}A_{m}
+\frac{1}{2}(\lambda_-\gamma^{m}\theta_-)D_{\alpha}A_{m}
+\frac{1}{2}(\gamma^m\lambda_-)_{\alpha}(\theta_-\gamma_m{\cal W})
\nn\\
&~~~~~~
+\frac{1}{4}(\gamma^m\theta_-)_{\alpha}(\lambda_-\gamma_m{\cal W})
+\frac{1}{16}(\gamma^m\theta_-)_{\alpha}(\gamma_m\theta_-)_{\gamma}\lambda_+^{\beta}\Fslash{\mathcal F}_{~\beta}^{\gamma}~,
\\
&\Theta^-_{\alpha} \equiv \frac{1}{2}\Bigl(c_1-\frac{1}{3} \Bigr)
(\gamma^m\theta_-)_{\alpha}(\lambda_+\gamma_m\theta_-)-\frac{1}{2}(\lambda_+\gamma^m\theta_-)(\gamma_m{\cal W})_{\alpha}~.
\end{align}
In the following,
we will examine conditions that each term in \bref{Total BRS surface}
vanishes.

To achieve our purpose, first, we focus on the term $\widetilde{\Pi}^i_{+}{\rm X}_i$ in (\ref{Total BRS surface}) which takes the form
\begin{align}
\widetilde{\Pi}^i_{+}\left[-\lambda_+^{\alpha}\gamma_{i\alpha\beta}\left(c_1\theta_-^{\beta}+{\cal W}^{\beta} \right)
+\delta_Q(y_i+A_i) \right]~.
\end{align}
Here we assume that
$\delta_Q(y_i+A_i)=0$.
This follows from the fact that
we fix degrees of freedom for the D-brane position by
\begin{align}
y_i=-A_i~.
\label{static gauge}
\end{align}
In fact,
this identification
turns to $y_i=0$ in the $\alpha'\to 0$ limit
after the scaling $A_i\rightarrow(2\pi\alpha')A_i$.
It implies that we consider a D-brane sitting at the origin.
As we will see below, the BRS transformation of (\ref{static gauge}) turns into one of the DBI equations and
the derivation of (\ref{static gauge}) respect to the time-coordinate $\tau$ also turns into the Dirichlet boundary condition.
One may add
a constant to the right hand side of \bref{static gauge}
to consider a D-brane sitting outside the origin,
but this will not affect the DBI equation
and the Dirichlet boundary condition
as anticipated.
In addition, we obtain the boundary condition on $\theta_-$ as
\begin{align}
\theta_-^{\beta}
=-\frac{1}{c_1}{\cal W}^{\beta}~.
\label{BC1DBI}
\end{align}
This eliminates $\theta_-^{\alpha}$ from (\ref{Total BRS surface}) completely.
Hereafter we understand $\theta_-^{\alpha}$ as (\ref{BC1DBI}).
Note that (\ref{BC1DBI}) also leads to
\begin{align}
\dot{\theta}_-^{\beta}=
-\frac{1}{c_1}
\left(\Pi_+^{\mu}\partial_{\mu}{\cal W}^{\beta}
+\dot{\theta}_+^{\gamma}D_{\gamma}{\cal W}^{\beta} \right)~.
\label{derivative of BC1}
\end{align}

Secondly, the terms $\Pi_-^i{\rm Y}_i$ and $\widetilde{\Pi}^{\mu}_{-}{\rm Y}_{\mu}$ reduce to
\begin{align}
\Pi_-^m\left[\left(c_2\lambda_-^{\alpha}+\frac{1}{4}\lambda_+^{\beta}\Fslash{\mathcal F}_{~\beta}^{\alpha} \right)\frac{1}{c_1}(\gamma_m{\cal W})_{\alpha} \right]~,
\end{align}
and imply  the boundary condition on $\lambda_-$
\begin{align}
\lambda_-^{\alpha}
=-\frac{1}{4c_2}\lambda_+^{\beta}
\,\,{\Fslash{\cal F}}_{~\beta}^{\alpha}~.
\label{BC2DBI}
\end{align}
This eliminates $\lambda_-^{\alpha}$ from (\ref{Total BRS surface}) completely.
Hereafter we understand $\lambda_-^{\alpha}$ as (\ref{BC2DBI}).

Here, it is better to comment on
two consequences of the boundary conditions (\ref{BC1DBI}) and (\ref{BC2DBI}).
First, consider the limit
$\alpha'\rightarrow 0$.
The limit $\alpha'\rightarrow 0$,
after rescaling 
$A_{\alpha}\rightarrow(2\pi\alpha')A_{\alpha}$, $A_m\rightarrow(2\pi\alpha')A_m$, 
${\cal W}^{\alpha}\rightarrow(2\pi\alpha'){\cal W}^{\alpha}$ and 
{${F}_{mn}\rightarrow(2\pi\alpha'){F}_{mn}$,}
turns
the boundary 
conditions (\ref{BC1DBI}) and (\ref{BC2DBI}) 
to the
usual boundary conditions $\theta_-^{\alpha}=\lambda_-^{\alpha}=0$.
The BRS invariance ${\delta}_Q(S_0+S_{\rm b}+V)=0$ 
then implies ${\delta}_QV=0$, 
since $\delta_Q(S_0+S_{\rm b})=0$ under these boundary conditions.
We can show that ${\delta}_QV=0$ with usual boundary conditions leads
to the super Yang-Mills equations of motion (\ref{SYM AB}), (\ref{SYM AW}) and (\ref{SYM WF})
as discussed in Appendix \ref{appendix:vertex}.

We  consider
the BRS variation $\delta_Q(y_i+A_i)=0$.
To evaluate it, we note that
$\delta_Q A_i=\lambda^{\alpha}_+D_{\alpha}A_i+\frac{1}{2}(\lambda_-\gamma^{\mu}\theta_-)\partial_{\mu}A_i$.
Under the boundary conditions (\ref{BC1DBI}) and (\ref{BC2DBI}), 
the equation $\delta_Q(y_i+A_i)=0$
is shown to reduce to the following equation
\begin{align}
-D_{\alpha}A_{i}
+\frac{1}{c_1}(\gamma_{i}{\cal W})_{\alpha}
-\frac{1}{8c_1c_2}\,\,{\Fslash{\cal F}}_{~\alpha}^{\beta}
(\gamma^{\mu}{\cal W})_{\beta}
\partial_{\mu}A_i
=0~.
\label{DBI-i}
\end{align}
This is one of the DBI equations.
Furthermore,
noting that $\dot A_i=\dot\theta_+^\alpha D_\alpha A_i +\Pi^\mu_+ \pa_\mu A_i$,
one finds that the time-derivative of \bref{static gauge}
turns out into the Dirichlet boundary condition given in \bref{Dirichlet BC DBI}.

Let us return to the subject. 
Thirdly, the term $\Delta^+_{\beta}\Lambda^{\beta}$ in (\ref{Total BRS surface}) is examined. We see that $\Lambda^{\beta}=0$ reduces to
\begin{align}
\frac{1}{c_1}D_{\alpha}{\cal W}^{\beta}
-\frac{1}{4c_2}\,\,{\Fslash{\cal F}}_{~\alpha}^{\beta}
+\frac{1}{8c_1^{\,2}c_2}\,\,{\Fslash{\cal F}}_{~\alpha}^{\gamma}
(\gamma^{\mu}{\cal W})_{\gamma}
\partial_{\mu}{\cal W}^{\beta}=0~.
\label{DBI-W}
\end{align}
This is one of the DBI equations on a D$p$-brane.
This equation ensures that 
conditions (\ref{BC1DBI}) and (\ref{BC2DBI}) are consistent with BRS transformations
$\delta_Q\theta_{-}^{\alpha}=\lambda_{-}^{\alpha}$ and $\delta_Q\lambda_{-}^{\alpha}=0$. 

As was done in  \cite{PS open background},
it is convenient to introduce a covariant derivative $\widehat D_\alpha$ 
by
\begin{align}
\widehat{D}_{\alpha}&\equiv
D_{\alpha}
+\frac{1}{8c_1c_2}\Fslash{\mathcal F}_{~\alpha}^{\gamma}
(\gamma^{\mu}{\cal W})_{\gamma}\partial_{\mu}~.
\end{align}
Applying it to $\frac{1}{c_1}\CW^\beta$, 
we obtain
\begin{align}
\frac{1}{c_1}\widehat{D}_{\alpha}\CW^\beta=
\frac{1}{c_1}D_{\alpha}\CW^\beta
+\frac{1}{8c_1^2c_2}\Fslash{\mathcal F}_{~\alpha}^{\gamma}
(\gamma^{\mu}{\cal W})_{\gamma}\partial_{\mu}\CW^\beta
=\frac{1}{4c_2}\Fslash{\mathcal F}_{~\alpha}^{\beta}~,
\label{BI Dirac type Eq}
\end{align}
where in the last equality \bref{DBI-W} is used.
On the other hand, as \bref{DBI-W} implies
\begin{align}
\frac{1}{4c_2}\Fslash{\mathcal F}_{~\alpha}^{\beta}
=\frac{1}{c_1}D_{\alpha}{\cal W}^{\gamma}\left(\delta_{\,\beta}^{\gamma}
-\frac{1}{2c_1^{\,2}}(\gamma^{\mu}{\cal W})_{\beta}\partial_{\mu}{\cal W}^{\gamma} \right)^{-1}~,
\end{align}
$\widehat D_\alpha$ is expressed as
\begin{eqnarray}
\begin{split}
\widehat{D}_{\alpha}
=D_{\alpha}+\frac{1}{2c_1^{\,2}}D_{\alpha}{\cal W}^{\delta}
\Bigl(\delta_{\,\gamma}^{\,\delta}-\frac{1}{2c_1^{\,2}}(\gamma^{\nu}{\cal W})_{\gamma}\partial_{\nu}{\cal W}^{\delta} \Bigr)^{-1}
(\gamma^{\mu}{\cal W})_{\gamma}\partial_{\mu}~.
\end{split}
\end{eqnarray}
It follows that it satisfies the following  anticommutation relation
\begin{eqnarray}
\begin{split}
&\{\widehat{D}_{\alpha},\,\widehat{D}_{\beta} \}
=\Bigl(\gamma^{\mu}_{\alpha\beta}+\frac{1}{16c_2^{\,2}}\Fslash{\mathcal F}_{~\alpha}^{\gamma}\Fslash{\mathcal F}_{~\beta}^{\delta}\gamma_{\gamma\delta}^{\mu} \Bigr)
\widehat{\partial}_{\mu}~,
\\
&\widehat{\partial}_{\mu}\equiv\partial_{\mu}+\frac{1}{2c_1^{\,2}}\partial_{\mu}{\cal W}^{\alpha}
\Bigl(\delta_{\,\beta}^{\,\alpha}-\frac{1}{2c_1^{\,2}}\gamma_{\beta\gamma}^{\nu}{\cal W}^{\gamma}\partial_{\nu}{\cal W}^{\alpha} \Bigr)^{-1}(\gamma^{\rho}{\cal W})_{\beta}\partial_{\rho}~.
\end{split}
\label{eqn:DD}
\end{eqnarray}

Fourthly, the term $(N_+)_{\beta}^{\,\alpha}{\rm Z}_{~\alpha}^{\beta}$ in (\ref{Total BRS surface}) is examined. 
Using (\ref{BI Dirac type Eq}) and \bref{eqn:DD}, it turns to
\begin{align}
\lambda^{\alpha}_+\lambda^{\gamma}_+
\Bigl(\frac{1}{c_2}
\widehat{D}_{\alpha}\,\,{\Fslash{\cal F}}^{\beta}_{~\gamma}\Bigr)
=\lambda^{\alpha}_+\lambda^{\gamma}_+
\Bigl(\frac{4}{c_1}\widehat{D}_{\alpha}\widehat{D}_{\gamma}{\cal W}^{\beta}\Bigr)
=\lambda^{\alpha}_+\lambda^{\gamma}_+
\Bigl(\gamma^{\mu}_{\alpha\gamma}+\frac{1}{16c_2^{\,2}}\Fslash{\mathcal F}_{~\alpha}^{\delta}\Fslash{\mathcal F}_{~\gamma}^{\eta}\gamma_{\delta\eta}^{\mu} \Bigr)
\frac{2}{c_1}\widehat{\partial}_{\mu}{\cal W}^{\beta}~,
\label{DBI-PS}
\end{align}
which vanishes due to the pure spinor constraint 
$\lambda_+\gamma^{\mu}\lambda_++\lambda_-\gamma^{\mu}\lambda_-=0$\,.

Fifthly, we consider terms including  $\Pi^{\mu}_{+}$ in (\ref{Total BRS surface}),
$\Pi_+^{\mu}{\rm X}_{\mu}-\frac{1}{c_1}\Pi^{\mu}_{+}\partial_{\mu}{\cal W}^{\alpha}\Theta_{\alpha}^-$, 
where the second term comes from (\ref{derivative of BC1}). 
It is straightforward to see that it is eliminated by
\begin{align}
&
\partial_{\mu}A_{\alpha}-D_{\alpha}A_{\mu}
+\frac{1}{c_1}(\gamma_{\mu}{\cal W})_{\alpha}
\nn\\&~~~
+\frac{1}{6c_1^{\,3}}(\gamma^{n}{\cal W})_{\alpha}
({\cal W}\gamma_{n}\partial_{\mu}{\cal W})
+\frac{1}{8c_1c_2}\,{\Fslash{\cal F}}_{~\alpha}^{\beta}
(\gamma^{n}{\cal W})_{\beta}
(\partial_{\mu}A_{n}-\partial_{n}A_{\mu})
=0~,
\label{DBI-mu}
\end{align}
which is one of the DBI equations.
Combining it with
(\ref{DBI-i}),
we obtain
\begin{align}
&
\partial_{m}A_{\alpha}-D_{\alpha}A_{m}
+\frac{1}{c_1}(\gamma_{m}{\cal W})_{\alpha}
\nn\\&~~~
+\frac{1}{6c_1^{\,3}}(\gamma^{n}{\cal W})_{\alpha}
({\cal W}\gamma_{n}\partial_{m}{\cal W})
+\frac{1}{8c_1c_2}\,{\Fslash{\cal F}}_{~\alpha}^{\beta}
(\gamma^{n}{\cal W})_{\beta}
(\partial_{m}A_{n}-\partial_{n}A_{m})
=0~.
\label{DBI-Am}
\end{align}

Finally, we consider terms including  $\dot{\theta}_+^{\alpha}$ in (\ref{Total BRS surface}),
$\dot{\theta}_+^{\alpha}\Theta^+_{\alpha}-\frac{1}{c_1}\dot{\theta}_+^{\beta}D_{\beta}{\cal W}^{\alpha}\Theta_{\alpha}^-$, 
where the second term comes from (\ref{derivative of BC1}). 
These terms are eliminated by
\begin{align}
&
-D_{\alpha}A_{\beta}-D_{\beta}A_{\alpha}
+\gamma^{m}_{\alpha\beta}A_{m}
-\frac{1}{6c_1^{\,3}}(\gamma^{m}{\cal W})_{\alpha}
({\cal W}\gamma_{m}D_{\beta}{\cal W})
\nn\\&~~
+\frac{1}{12c_1^{\,2}c_2}\,\,{\Fslash{\cal F}}_{~\alpha}^{\gamma}
(\gamma^{m}{\cal W})_{\beta}(\gamma_{m}{\cal W})_{\gamma}
-\frac{1}{8c_1c_2}\,\,{\Fslash{\cal F}}_{~\alpha}^{\gamma}
(\gamma^{m}{\cal W})_{\gamma}(\partial_mA_{\beta}-D_{\beta}A_m)=0~.
\end{align}
By eliminating $\partial_mA_{\beta}-D_{\beta}A_m$ by \bref{DBI-Am}, 
it reduces to 
\begin{align}
&
-D_{\alpha}A_{\beta}-D_{\beta}A_{\alpha}
+\gamma^{m}_{\alpha\beta}A_{m}
-\frac{1}{6c_1^{\,3}}(\gamma^{m}{\cal W})_{\alpha}
({\cal W}\gamma_{m}D_{\beta}{\cal W})
\nn\\&~~~
+\frac{1}{6c_1^{\,2}}(\gamma^{m}{\cal W})_{\beta}(\gamma_{m}{\cal W})_{\gamma}
\left\{-\frac{1}{4c_2}\,\,{\Fslash{\cal F}}_{~\alpha}^{\gamma}
+\frac{1}{8c_1^{\,2}c_2}\,\,{\Fslash{\cal F}}_{~\alpha}^{\delta}
(\gamma^{n}{\cal W})_{\delta}
\partial_{n}{\cal W}^{\gamma} \right\}
\nn\\
&~~~~~~+\frac{1}{64c_1^{\,2}c_2^{\,2}}
\,\,{\Fslash{\cal F}}_{~\alpha}^{\gamma}\,\,{\Fslash{\cal F}}_{~\beta}^{\delta}
(\gamma^{m}{\cal W})_{\gamma}(\gamma^{n}{\cal W})_{\delta}
(\partial_{m}A_{n}-\partial_{n}A_{m})
=0~.
\label{A_{alpha beta}}
\end{align}
Finally substituting \bref{DBI-W} into the expression in the curly braces
in \bref{A_{alpha beta}},
we obtain
\begin{align}
&
-D_{\alpha}A_{\beta}-D_{\beta}A_{\alpha}
+\gamma^{m}_{\alpha\beta}A_{m}
-\frac{1}{6c_1^{\,3}}(\gamma^{m}{\cal W})_{\alpha}
({\cal W}\gamma_{m}D_{\beta}{\cal W})
-\frac{1}{6c_1^{\,3}}(\gamma^{m}{\cal W})_{\beta}
({\cal W}\gamma_{m}D_{\alpha}{\cal W})
\nn\\
&~~~+\frac{1}{64c_1^{\,2}c_2^{\,2}}
\,\,{\Fslash{\cal F}}_{~\alpha}^{\gamma}\,\,{\Fslash{\cal F}}_{~\beta}^{\delta}
(\gamma^{m}{\cal W})_{\gamma}
(\gamma^{n}{\cal W})_{\delta}
(\partial_{m}A_{n}-\partial_{n}A_{m})=0~.
\label{DBI-alpha}
\end{align}

As a result, 
we have obtained not only boundary  conditions (\ref{BC1DBI}) and (\ref{BC2DBI}),
but also independent equations for background superfields 
\bref{DBI-Am}, (\ref{DBI-alpha}) and (\ref{DBI-W}) which eliminates (\ref{Total BRS surface}).
We note that $c_1$ and $c_2$ can be absorbed into  redefinitions of ${\cal W}^{\alpha}$ and $\ \displaystyle{\Fslash{\mathcal F}_{~\alpha}^{\beta}}$ as
$\frac{1}{c_1}{\cal W}^{\alpha} \rightarrow {\cal W}^{\alpha}$ and 
$\frac{1}{c_2} \ \displaystyle{\Fslash{\mathcal F}_{~\alpha}^{\beta}} \rightarrow \,\ \displaystyle{\Fslash{\mathcal F}_{~\alpha}^{\beta}}$.
So we will set $c_1=c_2=1$ without loss of generality\footnote
{If we construct the $\kappa$-invariant boundary term which cancels out an ${\cal N}=1$ supersymmetry variation of the Green-Schwarz action
 and turn it into the BRS-invariant boundary term like (\ref{SUSY counter term}) by the method used in \cite{OT} (see also the section 4.1 in \cite{ICTP}), it must be shown $c_1=c_2=1$. }. 

Summarizing, we have obtained supersymmetric DBI equations of motion on a D$p$-brane
\begin{align}
&
\partial_{m}A_{\alpha}-D_{\alpha}A_{m}
+(\gamma_{m}{\cal W})_{\alpha}
\nn\\&~~~~
+\frac{1}{6}(\gamma^{n}{\cal W})_{\alpha}
({\cal W}\gamma_{n}\partial_{m}{\cal W})
+\frac{1}{8}\,{\Fslash{\cal F}}_{~\alpha}^{\beta}
(\gamma^{n}{\cal W})_{\beta}
(\partial_{m}A_{n}-\partial_{n}A_{m})
=0~,
\label{BI1}\\ 
&
D_{\alpha}A_{\beta}+D_{\beta}A_{\alpha}
-\gamma^m_{\alpha\beta}A_m
+\frac{1}{6}(\gamma^m{\cal W})_{\alpha}
({\cal W}\gamma_mD_{\beta}{\cal W})
+\frac{1}{6}(\gamma^m{\cal W})_{\beta}
({\cal W}\gamma_mD_{\alpha}{\cal W})~~~~
\nn \\&
\hspace{50mm}
-\frac{1}{64}
\Fslash{\mathcal F}_{~\alpha}^{\gamma}\Fslash{\mathcal F}_{~\beta}^{\delta}
(\gamma^m{\cal W})_{\gamma}
(\gamma^n{\cal W})_{\delta}
(\partial_mA_n-\partial_nA_m)=0~,
\label{BI2} 
\\&
D_{\alpha}{\cal W}^{\beta}
-\frac{1}{4}\Fslash{\mathcal F}_{~\alpha}^{\beta}
+\frac{1}{8}\Fslash{\mathcal F}_{~\alpha}^{\gamma}
(\gamma^{\mu}{\cal W})_{\gamma}
\partial_{\mu}{\cal W}^{\beta}=0~.
\label{BI3}
\end{align}
In the last equation, the index $\mu$ may be replaced with $m$ because $\pa_i\CW^\beta=0$.
Now it is manifest that our DBI equations on a D$p$-brane 
can be expressed in a ten-dimensional covariant fashion.
In other words,
our result coincides with the dimensional reduction of those for a D9-brane,
though the ten-dimensional covariance was absent in the beginning
of our analysis.

\section{Summary and discussions \label{section 5}}
We have examined the BRS invariance of the open pure spinor superstring in the presence of background superfields
on a D$p$-brane. 
It was shown that the BRS invariance leads not only  to boundary conditions on the spacetime spinors,
but also to supersymmetric DBI equations of motion for the background superfields on a D$p$-brane.
These DBI equations precisely coincide with those obtained by a dimensional reduction of
the supersymmetric DBI equations for the abelian D9-brane given in \cite{PS open background, Ker D9 DBI}.

We have introduced the boundary term $S_\mathrm{b}$
and the background coupling $V$.
Both are determined by the BRS symmetry.
In fact, $S_\mathrm{b}$ was shown to satisfy $\delta_Q(S_0+S_\mathrm{b})=0$,
when we take the limit $\alpha'\to 0$ and turn off the background couplings.
As for $V$, we have shown that the conditions for $\delta_Q(S_0+S_\mathrm{b}+V)=0$
reduce to the dimensional reduction of the super-Yang-Mills equations when $\alpha'\to 0$.
In fact, taking the limit $\alpha' \rightarrow 0$, 
after rescaling 
$A_{\alpha}\rightarrow(2\pi\alpha')A_{\alpha}$, $A_m\rightarrow(2\pi\alpha')A_m$, 
${\cal W}^{\alpha}\rightarrow(2\pi\alpha'){\cal W}^{\alpha}$ and 
${F}_{mn}\rightarrow(2\pi\alpha'){F}_{mn}$,
the DBI equations (\ref{BI1})-(\ref{BI3}) reduce to the super Yang-Mills equations of motion (\ref{SYM AB})-(\ref{SYM WF})
with an appropriate dimensional reduction.

We note that the ten-dimensional Lorentz covariance
is manifestly broken by the boundary term $S_{\rm b}$ as well as the background coupling $V$.
However the obtained DBI equations can be expressed in a covariant form.
This implies that our result is consistent with that for a D9-brane.

We  expect that we can extend our result 
so that
the BRS invariance  should lead to supersymmetric non-abelian DBI equations of motion  on a D$p$-brane.
We would like to report this issue in the near future \cite{HS4}.

As an alternative to our study, non-abelian deformations of the maximally supersymmetric Yang-Mills theory can be specified based on spinorial cohomology \cite{spinorial cohomology}, which may be closely related to the pure spinor fields in ten- and eleven-dimensional spacetime \cite{N=10 SYM const,How line,How function}.
The structure of higher-derivative invariants in the maximally supersymmetric Yang-Mills theories are studied in \cite{HLW SC}.
Moreover, in \cite{Ced PS field,Ced overview} the pure spinor superspace formalism is developed, which contains not only (minimal) pure spinor variables but also non-minimal pure spinor variables \cite{NMPS}. 
This enables us to construct the BRS invariant action for the ten-dimensional supersymmetric DBI theory.
Recently, this off-shell action is studied further in \cite{MSYM2,Ber Cederwall action}.
It is interesting to pursue these issues from the open string point of view.

\medskip
 
On the other hand,  the classical BRS invariance of a closed pure spinor superstring
in a curved background is shown to imply that the background fields satisfy full non-linear equations of motion
for the type II supergravity \cite{PS closed background}. 
This is similar to the result for the classical $\kappa$-invariance of  a closed Green-Schwarz  
superstring \cite{GS closed background}.
Moreover, recently in \cite{TseWul SSE} the classical $\kappa$-invariance also leads to
 the generalized type II supergravity equations of motion\footnote{
See \cite{GSEDFT} for further investigations based on double field theory.
} 
whose solutions originally have been found out in the context of integrable deformations of \adss{5}{5} sigma models \cite{generalized SUGRA}.
It is also interesting to consider whether the generalization of DBI equations can be derived analogously
from the $\kappa$- or BRS-invariance of an open superstring.

An immediate task is to clarify contribution of the dilaton superfield to Bianchi identities.
In that case we need to investigate closely the DBI equation corresponding to $I_{mn\alpha}=0$ in the super Yang-Mills theory as we see in Appendix \ref{appendix:SYM}.
This equation is also useful to confirm that our result agrees with the one which comes from the bosonic part of the DBI action.

Finally, it is interesting for us to calculate quantum higher-derivative corrections to our result 
by analyzing the quantum BRS invariance of the open pure spinor superstring.

\section*{Acknowledgments}

The authors would like to thank Takanori Fujiwara, Yoshifumi Hyakutake and Kentaroh Yoshida
 for useful comments. 
SH also thanks the Yukawa Institute for Theoretical Physics at Kyoto University.
Discussions during the workshop YITP-T-18-04 ``New Frontiers in String Theory 2018''
were useful to complete this work.
HS and MS would like to appreciate the organizers of the conferences
``KEK Theory Workshop 2018" held at KEK theory center, Tsukuba, December 17 - 20, 
2018,
and
The 2nd Workshop on ``Mathematics and Physics in General Relativity",
held at Setsunan University,
March 23 - 24, 2019,
for their kind hospitality. 
There, HS and MS reported the main results in this paper. 
 
\appendix

\section*{Appendix}

\section{Ten-dimensional ${\mathcal N}=1$ super Yang-Mills space \label{appendix:SYM}}

We will review the ten-dimensional $\CN=1$
 super-Yang-Mills theory \cite{Wit Sie SYM}.
 Introducing a superconnection one-form $A=E^MA_M$, where $E^M$ are supervielbeins and $A_M=(A_m,A_{\alpha})$ are superconnections, 
we define the gauge supercovariant derivative $\nabla_M$
\begin{align}
\nabla_m=\partial_m+A_m~,~~~~\nabla_{\alpha}=D_{\alpha}+A_{\alpha}~.
\end{align}
where
$D_{\alpha}$ is the supercovariant derivative
defined by
\begin{align}
D_{\alpha}=\frac{\partial}{\partial\theta^{\alpha}}+\frac{1}{2}(\gamma^m\theta)_{\alpha}\partial_m~,
\label{covariant d}
\end{align}
which satisfies $\{D_{\alpha},\,D_{\beta}\}=\gamma^m_{\alpha\beta}\partial_m$. 
The field strengths $F_{MN}$ are defined by
\begin{align}
\left[\nabla_M,\,\nabla_N \right\}={T_{MN}}^R\nabla_R+{F}_{MN}~,
\label{def F 10d SYM}
\end{align}
where ${T_{MN}}^R$ are flat torsion tensors whose components
are fixed to zero except for $T_{\alpha\beta}^m=\gamma_{\alpha\beta}^m$. 
According to this definition, these field strengths are invariant under the gauge transformations
with a superfield parameter $\Omega$
\begin{align}
\delta A_m=\partial_m\Omega~,~~~~\delta A_{\alpha}=D_{\alpha}\Omega~.
\end{align}
For the on-shell super Yang-Mills theory, we might adopt a constraint \cite{N=10 SYM const} (see also \cite{How line})
\begin{align}
{F}_{\alpha\beta}=0~,
\label{constraint in 10d SYM}
\end{align}
which implies
\begin{align}
D_{\alpha}A_{\beta}+D_{\beta}A_{\alpha}+\{A_{\alpha},\,A_{\beta} \}
=\gamma^m_{\alpha\beta}A_m~.
\label{SYM1}
\end{align}
If we consider a dimensional reduction
to four-dimensions, 
we see that this constraint reduces to the one
 in the four-dimensional ${\cal N}=4$ super Yang-Mills theory \cite{N=4 SYM const}. 
 
In the following, let us solve the Bianchi identities represented as
\begin{align}
I_{MNR}=&(-1)^{R(M+N)}\nabla_R{F}_{MN}-{T_{MN}}^S{F}_{SR}
\nonumber \\
&+(-1)^{M(R+N)}\nabla_N{F}_{RM}-(-1)^{R(M+N)}{T_{RM}}^S{F}_{SN}
\nonumber \\
&+\nabla_M{F}_{NR}-(-1)^{M(N+R)}{T_{NR}}^S{F}_{SM}~.
\label{Bianchi id}
\end{align}
The first identity $I_{\alpha\beta\gamma}=0$ implies
\begin{align}
-\gamma_{\alpha\beta}^m{F}_{m\gamma}-\gamma_{\gamma\alpha}^m{F}_{m\beta}
-\gamma_{\beta\gamma}^m{F}_{m\alpha}=0~.
\end{align}
Thanks to the Fierz identity, we find that the field strength $F_{m\alpha}$
 must take the form of 
\begin{align}
{F}_{m\alpha}=-\gamma_{m\alpha\beta}{\cal W}^{\beta}~.
\label{def W in 10d SYM}
\end{align}
In other words,
\begin{align}
\partial_mA_{\alpha}-D_{\alpha}A_m+[A_m,\,A_{\alpha}]=-\gamma_{m\alpha\beta}{\cal W}^{\beta}~.
\label{SYM2}
\end{align}
Next the second identity $I_{m\alpha\beta}=0$ together (\ref{def W in 10d SYM}) implies
\begin{align}
\gamma_{m\alpha\delta}\nabla_{\beta}{\cal W}^{\delta}
+\gamma_{m\beta\delta}\nabla_{\alpha}{\cal W}^{\delta}
-\gamma_{\alpha\beta}^n{F}_{nm}=0~.
\end{align}
Multiplying this by $\gamma^{\alpha\beta}_p$, we find that
\begin{align}
{F}_{mn}=\frac{1}{8}(\gamma_{mn})_{\alpha}^{~\beta}\nabla_{\beta}{\cal W}^{\alpha}~,
\end{align}
which is equivalent to
\begin{align}
\nabla_{\alpha}{\cal W}^{\beta}=-\frac{1}{4}(\gamma^{mn})_{\alpha}^{~\beta}{F}_{mn}~.
\label{SYM3}
\end{align}
The third identity $I_{mn\alpha}=0$ implies
\begin{align}
\nabla_{\alpha}{F}_{mn}
=\gamma_{n\alpha\beta}\nabla_m{\cal W}^{\beta}
-\gamma_{m\alpha\beta}\nabla_n{\cal W}^{\beta}~.
\label{SYM4}
\end{align}
Taking (\ref{SYM3}) into account, (\ref{SYM4}) yields the result
\begin{align}
\gamma_{\alpha\beta}^m\nabla_m{\cal W}^{\beta}=0~.
\label{Dirac 10d W}
\end{align}
Furthermore, multiplying (\ref{Dirac 10d W}) by $\gamma^{n\gamma\alpha}\nabla_{\gamma}$ we find
\begin{align}
\nabla^m{F}_{mn}=-\frac{1}{2}\gamma_{n\alpha\beta}\left\{{\cal W}^{\alpha},\,{\cal W}^{\beta} \right\}~.
\label{Dirac 10d F}
\end{align}
The (\ref{Dirac 10d F}) and (\ref{Dirac 10d W})
 imply the Maxwell equation for the gauge field $\nabla_mf^{mn}=0$
and the Dirac equation for the gaugino $\gamma_{\alpha\beta}^m\nabla_m\xi^{\beta}=0$, respectively.

Finally, the remaining identity $I_{mnp}=0$ implies
\begin{align}
\nabla_m{F}_{np}+\nabla_n{F}_{pm}+\nabla_p{F}_{mn}=0~,
\end{align}
and it suggests that $F_{mn}$ is just the curl of a gauge field $A_m$; 
\begin{align}
{F}_{mn}=\partial_mA_n-\partial_nA_m+[A_m,\,A_n]~.
\label{SYM5}
\end{align}
The $\theta$-expansion of these superfields are studied in \cite{theta expansion}.


\section{Massless vertex operator for pure spinor open superstring \label{appendix:vertex}}

We present a review of  the vertex operators in the open pure spinor superstring \cite{PS}
 (see also \cite{ICTP}).
For simplicity, we focus on the left-moving sector only.

We consider a ghost number 1 massless vertex operator given by
\begin{align}
U=\lambda^{\alpha}A_{\alpha}(x,\theta)~,
\end{align}
where $A_{\alpha}(x,\theta)$ is a spinor superfield. 
The BRS transformation law is represented as
\begin{align}
Qx^m=\frac{1}{2}\lambda\gamma^m\theta~,~~~Q\theta^{\alpha}=\lambda^{\alpha}~,~~~
Qd_{\alpha}=-\Pi^m(\gamma_m\lambda)_{\alpha}~,~~~
Q\lambda^{\alpha}=0~,~~~Q\omega_{\alpha}=d_{\alpha}~,
\end{align}
where $Q$ denotes $\delta_1$ in section \ref{sec:BRS}.
Note that $Q^2\omega_{\alpha}=-\Pi^m(\gamma_m\lambda)_{\alpha}$ turns out the gauge transformation for $\omega_{\alpha}$.
Then cohomology condition,
$QU=0$ up to the gauge transformation ${\delta}U=Q\Omega$,
implies
\begin{align}
D_{\alpha}(\gamma_{mnpqr})^{\alpha\beta}A_{\beta}=0~~~~\mbox{and}~~~~
{\delta}A_{\alpha}=D_{\alpha}\Omega~,
\label{gauge conditions}
\end{align}
where $\Omega(x,\theta)$ is a gauge parameter and 
the derivative $D_{\alpha}$ is given in \bref{covariant d}.

To derive \bref{gauge conditions}, 
we use 
the pure spinor constraint
for the commutative bispinor $\lambda$
\begin{align}
\lambda^{\alpha}\lambda^{\beta}
=\frac{1}{2^5\,5!}\gamma^{\alpha\beta}_{mnpqr}\left(\lambda^{\gamma}\gamma_{\gamma\delta}^{mnpqr}\lambda^{\delta}\right)~.
\end{align}
As a result, \bref{gauge conditions}
is
consistent with the super Yang-Mills equations of motion 
and the gauge transformations as we have seen in Appendix A.

Next, we derive an integrated vertex operator such as $V=\int dz\,{\cal V}$. 
Recalling the RNS formulation, ${\cal V}$ is given as the anticommutator of the unintegrated vertex operator $U$ and the $b$-ghost. 
However, in the pure spinor formulation, the reparametrization $b$-ghost is unclear without introducing the non-minimal part \cite{NMPS}\footnote
{The non-minimal pure spinor formalism extended to the Maxwell background is investigated in \cite{Bak b}. }.
Fortunately, the above facts can be rephrased in terms of the BRS charge $Q$ as\,\footnote
{The Jacobi identity implies
\begin{align}
Q{\cal V}=\bigl[Q,\,\bigl\{\oint dz\,b,\,U \bigr\}\bigr]=
-\bigl[U,\,\bigl\{Q,\,\oint dz\,b\bigr\} \bigr]-\big[\oint dz\,b,\,\bigl\{U,\,Q \bigr\} \bigr]={\partial}U
\end{align}
since $\{Q,\,U \}=0$, $\{Q,\,b \}=T$ and $[\oint dz\,T,\,U]=\partial{U}$ for the conformal weight zero primary operator $U$.}
\begin{align}
Q{\cal V}=\partial U~.
\end{align}
We find the vertex operator ${\cal V}$ takes the form of
\begin{align}
{\cal V}=\partial\theta^{\alpha}A_{\alpha}(x,\theta)+\Pi^mA_m(x,\theta)+d_{\alpha}{\cal W}^{\alpha}(x,\theta)+\frac{1}{2}N^{mn}{F}_{mn}(x,\theta)~,
\label{vertex}
\end{align}
where $N^{mn}=\frac{1}{2}\lambda\gamma^{mn}\omega$ is the ghost Lorentz current. 
Indeed, 
since
\begin{align}
Q{\cal V}&=\partial(\lambda^{\alpha}A_{\alpha})+\lambda^{\alpha}\partial\theta^{\beta}(-D_{\alpha}A_{\beta}-D_{\beta}A_{\alpha}+\gamma^m_{\alpha\beta}A_m)
\nonumber \\
&+\,\lambda^{\alpha}\Pi^m(D_{\alpha}A_m-\partial_mA_{\alpha}-\gamma_{m\alpha\beta}{\cal W}^{\beta})
\nonumber \\
&+\,\lambda^{\alpha}d_{\beta}\Bigl(-D_{\alpha}{\cal W}^{\beta}+\frac{1}{4}(\gamma^{mn})_{\alpha}^{~\beta}{F}_{mn} \Bigr)+\frac{1}{2}\lambda^{\alpha}N^{mn}D_{\alpha}{F}_{mn}~,
\end{align}
\bref{vertex} implies 
the following equations
\begin{align}
-D_{\alpha}A_{\beta}-D_{\beta}A_{\alpha}+\gamma^m_{\alpha\beta}A_m&=0~,
\label{SYM AB} \\
D_{\alpha}A_m-\partial_mA_{\alpha}-\gamma_{m\alpha\beta}{\cal W}^{\beta}&=0~,
\label{SYM AW} \\
-D_{\alpha}{\cal W}^{\beta}+\frac{1}{4}(\gamma^{mn})_{\alpha}^{~\beta}{F}_{mn}&=0~,
\label{SYM WF} \\
\lambda^{\alpha}\lambda^{\beta}(\gamma^{mn})_{\beta}^{~\gamma}D_{\alpha}{F}_{mn}&=0~.
\label{SYM PS}
\end{align}
The 
(\ref{SYM AB}), (\ref{SYM AW}) and (\ref{SYM WF}) certainly correspond to 
the super-Yang-Mills equations
(\ref{SYM1}), (\ref{SYM2}) and (\ref{SYM3}) in the abelian case, respectively.
It follows that superfields $A_{\alpha}$ and $A_m$ are spinor and vector gauge fields in the ten-dimensional ${\cal N}=1$ super Yang-Mills theory,
and that ${\cal W}^{\alpha}$ and ${F}_{mn}$ are spinor and vector field strengths for them. 
On the other hand, (\ref{SYM PS}) 
is satisfied by the pure spinor constraint
\begin{align}
\lambda^{\alpha}\lambda^{\beta}(\gamma^{mn})_{\beta}^{~\gamma}D_{\alpha}{F}_{mn}
=4\lambda^{\alpha}\lambda^{\beta}D_{\alpha}D_{\beta}{\cal W}^{\gamma}
=2(\lambda\gamma^m\lambda)\partial_m{\cal W}^{\gamma}=0~,
\end{align}
where  \bref{SYM WF} is used.
If (\ref{SYM AB}) is contracted with $(\gamma_{mnpqr})^{\alpha\beta}$, we obtain the equation of motion for $A_{\alpha}$ in (\ref{gauge conditions}). 
Contraction of (\ref{SYM AB}) with $\gamma_n^{\alpha\beta}$ also leads to
\begin{align}
A_m=\frac{1}{8}\gamma_m^{\alpha\beta}D_{\alpha}A_{\beta}~.
\label{SYM A}
\end{align}
Then the gauge transformation in (\ref{gauge conditions}) turns to
$\delta A_m=\partial_m\Omega$. 
Similarly contracting (\ref{SYM AW}) with $\gamma^{m\alpha\gamma}$ implies  the equation for ${\cal W}^{\alpha}$
\begin{align}
{\cal W}^{\beta}=\frac{1}{10}\gamma^{m\alpha\beta}\left(D_{\alpha}A_m-\partial_mA_{\alpha}\right)~,
\label{SYM W}
\end{align}
and contracting (\ref{SYM WF}) with $(\gamma^{pq})_{\beta}^{\,\alpha}$ implies the equation for ${F}_{mn}$
\begin{align}
{F}_{mn}=\frac{1}{8}{(\gamma_{mn})_{\alpha}}^{\beta}D_{\beta}{\cal W}^{\alpha}~.
\label{SYM F}
\end{align}

Furthermore, utilizing (\ref{SYM A}), (\ref{SYM AW}) and (\ref{SYM F}), we derive
\begin{align}
\partial_{[m}A_{n]}&
=-\frac{1}{8}\gamma_{[m}^{\alpha\beta}D_{\alpha}(\partial_{n]}A_{\beta})
=-\frac{1}{8}\gamma_{[m}^{\alpha\beta}D_{\alpha}\left(D_{\beta}A_{n]}-(\gamma_{n]}{\cal W})_{\beta} \right)
\nn \\
&=\frac{1}{8}(\gamma_{mn})_{\beta}^{\,\alpha}D_{\alpha}{\cal W}^{\beta}={F}_{mn}~.
\label{SYM FA}
\end{align}
Besides, this equation together (\ref{SYM AW}) implies 
\begin{align}
D_{\alpha}{F}_{mn}=\partial_{[m}D_{|\alpha|}A_{n]}=\partial_{[m}(\gamma_{n]}{\cal W})_{\alpha}~.
\label{SYM FW}
\end{align}
(\ref{SYM FA}) and (\ref{SYM FW}) certainly correspond to remaining Bianchi identities (\ref{SYM5}) and (\ref{SYM4}) for the abelian case, respectively.


\section{BRS charge conservation} \label{appendix:charge conservation}

We will derive the supersymmetric DBI equations 
by modifying the method used in \cite{PS open background}
 to include the Dirichlet components.

We require that the general variation $\delta(S_0+S_{\rm b}+V)$ vanishes. 
This leads to boundary conditions in the presence of background superfields. 
Under these conditions, it is shown that the BRS charge conservation implies superfield equations for DBI fields.

Let us begin to examine a general variation of the world-sheet action $S_0$ in \bref{PS action}, its ten-dimensional ${\mathcal N}=1$ supersymmetry counter-term $S_{\rm b}$
in \bref{SUSY counter term} and the background coupling $V$ in \bref{background action}. 
We find that variations $\delta(S_0+S_{\rm b})$ and $\delta{V}$ may be expressed as
\begin{align}
{\delta}(S_0+S_{\rm b})=\frac{1}{2\pi\alpha'}&\int{\rm d}\tau\biggl\{\delta\theta_+^{\alpha}
\Bigl[\, \frac{1}{2}d^-_{\alpha}+\Pi_+^{\mu}(\gamma_{\mu}\theta_-)_{\alpha}
+\widetilde{\Pi}_{+}^{i }(\gamma_{i }\theta_-)_{\alpha}-y_i(\gamma^i\dot{\theta}_+)_{\alpha}
\nn\\
&~~ 
+\frac{1}{6}(\theta_-\gamma^m\dot{\theta}_-)(\gamma_m\theta_-)_{\alpha}
\Bigr]
+\delta\theta_-^{\alpha}\Bigl[
\,\frac{1}{2}\left(1-c_1 \right)\Delta^+_{\alpha}
-\frac{1}{6}(\theta_-\gamma^m\dot{\theta}_+)(\gamma_m\theta_-)_{\alpha}
\Bigr]
\nn\\
&
-{\delta}y_+^{\mu}\Bigl[
\widetilde{\Pi}_{-\mu}-\frac{1}{2}(\theta_-\gamma_{\mu}\dot{\theta}_+) \Bigr]
+{\delta}\widetilde{\Pi}_+^iy_i
+\frac{1}{2}c_1\delta\Delta^+_{\alpha}\theta_-^{\alpha}
\nn\\
&
+\frac{1}{2}\left(c_2-1 \right)\omega^+_{\alpha}\delta\lambda_-^{\alpha}+\frac{1}{2}c_2\delta\omega^+_{\alpha}\lambda_-^{\alpha}-\frac{1}{2}\omega^-_{\alpha}\delta\lambda_+^{\alpha}
\biggr\}~,
\label{delta 0b}
\end{align}
\begin{align}
{\delta}V=\frac{1}{2\pi\alpha'}&\int{\rm d}\tau\biggl\{\delta\theta_+^{\alpha}
\Bigl[\dot{\theta}_+^{\beta}
(\gamma_{\alpha\beta}^{\mu}A_{\mu}-D_{\alpha}A_{\beta}-D_{\beta}A_{\alpha})
+\Pi_+^{\mu}(D_{\alpha}A_{\mu}-\partial_{\mu}A_{\alpha})+\widetilde{\Pi}_{+}^{i }D_{\alpha}A_i
\Bigr.\biggr.\nn\\
&~~
-\frac{1}{2}\Delta_{\beta}^+D_{\alpha}{\cal W}^{\beta}+\frac{1}{4}D_{\alpha}(N_+\Fslash{\mathcal F})\Bigr]
+{\delta}y_{+}^{\mu}\Bigl[\dot{\theta}_+^{\alpha}
(\partial_{\mu}A_{\alpha}-D_{\alpha}A_{\mu})
+\Pi_+^{\nu}(\partial_{\mu}A_{\nu}-\partial_{\nu}A_{\mu})
\nn\\
&~~+\widetilde{\Pi}_{+}^{i }\partial_{\mu}A_i 
+\frac{1}{2}\Delta_{\alpha}^+\partial_{\mu}{\cal W}^{\alpha}+\frac{1}{4}\partial_{\mu}(N_+\Fslash{\mathcal F}) \Bigr]
+{\delta}\widetilde{\Pi}_+^iA_i
\nn\\
&+\frac{1}{2}{\delta}\Delta_{\alpha}^+{\cal W}^{\alpha}
+\frac{1}{8}{\delta}\lambda_+^{\alpha}\Fslash{\mathcal F}_{~\alpha}^{\beta}\omega_{\beta}^+
+\frac{1}{8}\lambda_+^{\alpha}\Fslash{\mathcal F}_{~\alpha}^{\beta}{\delta}\omega_{\beta}^+\biggl.\biggr\}~,
\label{delta v}
\end{align}
where $\delta y^\mu$ defined by
\begin{align}
\delta y_+^{\mu}=\delta x^{\mu}+\frac{1}{2}(\theta_+\gamma^{\mu}\delta\theta_+)~
\end{align}
is invariant under the ${\cal N}=1$ supersymmetry. 
We also see that ${\delta}(S_0+S_{\rm b})/{\delta y_i}\Bigr|=0$ as mentioned in section 3.

To obtain boundary conditions from
$\delta(S_0+S_{\rm b}+V)=0$, first we focus on the terms with $\delta\Delta^+_{\alpha}$ and $\delta\omega^+_{\alpha}$, and derive
\begin{align}
&\theta_-^{\alpha}
=-\frac{1}{c_1}{\cal W}^{\alpha}~,~~~~
\lambda_-^{\alpha}
=-\frac{1}{4c_2}\lambda_+^{\beta}
\,\,{\Fslash{\cal F}}_{~\beta}^{\alpha}~.
\label{BC1'DBI}
\end{align}
They also lead to
\begin{eqnarray}
\begin{split}
&\dot{\theta}_-^{\alpha}=
-\frac{1}{c_1}
\left(\Pi_+^{\mu}\partial_{\mu}{\cal W}^{\alpha}
+\dot{\theta}_+^{\beta}D_{\beta}{\cal W}^{\alpha} \right)~,~~~
\delta{\theta}_-^{\alpha}
=-\frac{1}{c_1}
\left(\delta y_+^{\mu}\partial_{\mu}{\cal W}^{\alpha}
+\delta\theta_+^{\beta}D_{\beta}{\cal W}^{\alpha} \right)~,
\\
&\delta\lambda_-^{\alpha}
=-\frac{1}{4c_2}\delta\lambda_+^{\beta}\Fslash{\mathcal F}^{\alpha}_{~\beta}
-\frac{1}{4c_2}\left(\lambda_+^{\beta}\delta y_+^{\mu}\partial_{\mu}\Fslash{\mathcal F}^{\alpha}_{~\beta}+\lambda_+^{\beta}\delta\theta_+^{\gamma}D_{\gamma}\Fslash{\mathcal F}^{\alpha}_{~\beta} \right)~.
\end{split}
\end{eqnarray}
Next, examining the terms with $\delta\lambda_+^{\alpha}$ in $\delta(S_0+S_{\rm b}+V)$ we find
\begin{align}
\omega^-_{\alpha}
=\frac{1}{4c_2}\omega^+_{\beta}
\,\,{\Fslash \cal F}_{~\alpha}^{\beta}~.
\label{BC1''DBI}
\end{align}
Boundary conditions for $\lambda_-^{\alpha}$ in (\ref{BC1'DBI}) and $\omega^-_{\alpha}$ in (\ref{BC1''DBI}) are consistent with the ghost number charge conservation 
$\lambda^{\alpha}\omega_{\alpha}\bigr|
=\widehat{\lambda}^{\alpha}\widehat{\omega}_{\alpha}\bigr|$,
where ``$\big|$'' means ``evaluated at the boundary''.
On the other hand, we can eliminate the terms with ${\delta}\widetilde{\Pi}_+^i$ in $\delta(S_0+S_{\rm b}+V)$ by the identification (\ref{static gauge}).
After substituting above conditions into $\delta(S_0+S_{\rm b}+V)=0$, 
we examine the terms with $\delta y_+^\mu$ and $\delta\theta_+^{\alpha}$.
They lead to complicated boundary conditions
\begin{align}
\widetilde{\Pi}_{-\mu}=&~
\dot{\theta}_+^{\alpha}\biggl(
\partial_{\mu}A_{\alpha}-D_{\alpha}A_{\mu}
+\frac{1}{2c_1}(\gamma_{\mu}{\cal W})_{\alpha}
+\frac{1}{6c_1^{\,3}}(\gamma^{m}{\cal W})_{\alpha}
({\cal W}\gamma_{m}\partial_{\mu}{\cal W})
\biggr)
\nn \\
&+\,\Pi_{+}^{\nu}\left(
\partial_{\mu}A_{\nu}-\partial_{\nu}A_{\mu}
\right)
+\widetilde{\Pi}_+^i\partial_{\mu}A_i
+\frac{1}{2c_1}
\Delta_{\alpha}^+\partial_{\mu}{\cal W}^{\alpha}
+\frac{1}{4c_2}\partial_{\mu}(N_+\Fslash{\mathcal F})~,
\label{BC2'DBI}
\\[0.2cm]
\frac{1}{2}d_{\alpha}^-=&~
\dot{\theta}_+^{\beta}\biggl(
D_{\alpha}A_{\beta}+D_{\beta}A_{\alpha}
-\gamma^{m}_{\alpha\beta}A_{m}
-\frac{1}{6c_1^{\,3}}
({\cal W}\gamma^mD_{\beta}{\cal W})
(\gamma_m{\cal W})_{\alpha}
-\frac{1}{6c_1^{\,3}}
({\cal W}\gamma^mD_{\alpha}{\cal W})
(\gamma_m{\cal W})_{\beta}
\biggr)
\nn \\
&+\Pi_{+}^{\mu}\biggl(
\partial_{\mu}A_{\alpha}-D_{\alpha}A_{\mu}
+\frac{1}{c_1}(\gamma_{\mu}{\cal W})_{\alpha}
+\frac{1}{6c_1^{\,3}}
({\cal W}\gamma^{m}\partial_{\mu}{\cal W})
(\gamma_{m}{\cal W})_{\alpha} \biggr)
\nn \\
&+\widetilde{\Pi}^i_{+}
\left(\frac{1}{c_1}(\gamma_i{\cal W})_{\alpha}-D_{\alpha}A_i \right)
+\frac{1}{2c_1}\Delta_{\beta}^+D_{\alpha}{\cal W}^{\beta}
-\frac{1}{4c_2}D_{\alpha}(N_+\Fslash{\mathcal F})~.
\label{BC3'DBI}
\end{align}
The (\ref{BC2'DBI}) is regarded as a 
modified Neumann boundary condition.
Boundary conditions for $\omega^-_{\alpha}$ in (\ref{BC1''DBI})
and $d^-_{\alpha}$ in (\ref{BC3'DBI}) must be consistent with the BRS transformation
$\delta_Q\omega^{-}_{\alpha}=d^{-}_{\alpha}$
up to the $\Lambda$-gauge transformation in section 2.
In the following discussion, 
we will absorb $c_1$ and $c_2$ by rescaling 
${\cal W}^{\alpha} \rightarrow c_1{\cal W}^{\alpha}$ and 
$\ \displaystyle{\Fslash{\mathcal F}^{\beta}_{~\alpha}} \rightarrow c_2\,\,\displaystyle{\Fslash{\mathcal F}^{\beta}_{~\alpha}}$.

To extract DBI equations,
we impose the following relation for BRS currents
\begin{align}
\lambda^{\alpha}d_{\alpha}\Bigr|
=\widehat{\lambda}^{\alpha}\widehat{d}_{\alpha}\Bigr|~,
\label{BRS currents}
\end{align}
which implies BRS charge conservation
\begin{align}
0&=\partial_{\tau}Q_{\rm total}=\int{\rm d}\sigma\,
\partial_{\tau}(j_{\rm BRS}^{\tau})
=\int{\rm d}\sigma\,
\partial_{\sigma}(j_{\rm BRS}^{\sigma})
\nn \\
&~~=\int{\rm d}\sigma\,
\partial_{\sigma}
(j_{\rm BRS}^z
-j_{\rm BRS}^{\bar{z}})
=\Bigl(\lambda^{\alpha}d_{\alpha}
-\widehat{\lambda}^{\alpha}\widehat{d}_{\alpha} \Bigr)
\Bigr|~.
\label{charge conservation}
\end{align}
Then we assume the Dirichlet boundary condition
\begin{align}
\Pi_{-i}=-\Pi_+^{\mu}\partial_{\mu}A_i-\dot{\theta}_+^{\alpha}D_{\alpha}A_i
+\frac{1}{2c_1}(\dot{\theta}_+\gamma_i{\cal W})~.
\label{Dirichlet BC DBI}
\end{align}
This is parallel with the Neumann boundary condition in (\ref{BC2'DBI}) and just the derivation of the identification (\ref{static gauge}) respect to the time-coordinate $\tau$.

Under these boundary conditions (\ref{BC1'DBI}), (\ref{BC2'DBI}), (\ref{BC3'DBI}) and (\ref{Dirichlet BC DBI}), the BRS charge conservation (\ref{BRS currents}) implies

\begin{align}
0&=\widehat{\lambda}^{\alpha}\widehat{d}_{\alpha}
-\lambda^{\alpha}d_{\alpha}
\nn \\
&=\frac{1}{2}\lambda_+^{\alpha}d^-_{\alpha}
+\frac{1}{2}\lambda_-^{\alpha}\Delta_{\alpha}^+
-\frac{1}{2}(\lambda_-\gamma_{\mu}\theta_-)\widetilde{\Pi}_{-}^{\mu}
-\frac{1}{2}(\lambda_-\gamma_i\theta_-)\Pi_{-}^i
-\frac{1}{4}(\lambda_-\gamma^m\theta_-)
(\theta_-\gamma_m\dot{\theta}_+)
\nn \\
&=\lambda^{\alpha}_+\dot{\theta}_+^{\beta}\Bigl[
D_{\alpha}A_{\beta}+D_{\beta}A_{\alpha}
-\gamma^m_{\alpha\beta}A_m
-\frac{1}{6}
({\cal W}\gamma^{m}D_{\beta}{\cal W})
(\gamma_{m}{\cal W})_{\alpha}
-\frac{1}{6}
({\cal W}\gamma^{m}D_{\alpha}{\cal W})
(\gamma_{m}{\cal W})_{\beta}
\nn \\
&~~~~+\frac{1}{8}\,\Fslash{\mathcal F}_{~\alpha}^{\gamma}
(\gamma^{m}{\cal W})_{\gamma}
\Bigl\{
\partial_{m}A_{\beta}-D_{\beta}A_{m}
+(\gamma_{m}{\cal W})_{\beta}
+\frac{1}{6}(\gamma^{n}{\cal W})_{\beta}
({\cal W}\gamma_{n}\partial_{m}{\cal W})
\Bigr\}\Bigr]
\nn \\
&+\lambda^{\alpha}_+
\Pi_{+}^{\mu}
\Bigl[
\partial_{\mu}A_{\alpha}-D_{\alpha}A_{\mu}
+(\gamma_{\mu}{\cal W})_{\alpha}
+\frac{1}{6}({\cal W}\gamma^{m}\partial_{\mu}{\cal W})
(\gamma_{m}{\cal W})_{\alpha}
+\frac{1}{8}\,\Fslash{\mathcal F}_{~\alpha}^{\beta}
(\gamma^{n}{\cal W})_{\beta}
(\partial_{\mu}A_{n}-\partial_{n}A_{\mu})
\Bigr]
\nn \\
&+\lambda^{\alpha}_+\widetilde{\Pi}_{+}^{i}
\Bigl[
-D_{\alpha}A_i+(\gamma_i{\cal W})_{\alpha}-\frac{1}{8}\,\Fslash{\mathcal F}_{~\alpha}^{\beta}(\gamma^{\mu}{\cal W})_{\beta}\partial_{\mu}A_i
\Bigr]
\nn \\
&+\frac{1}{2}\lambda^{\alpha}_+\Delta_{\beta}^+
\Bigl[
D_{\alpha}{\cal W}^{\beta}
-\frac{1}{4}\,\Fslash{\mathcal F}_{~\alpha}^{\beta}
+\frac{1}{8}\,\Fslash{\mathcal F}_{~\alpha}^{\gamma}
(\gamma^{\mu}{\cal W})_{\gamma}
\partial_{\mu}{\cal W}^{\beta}
\Bigr]
\nn \\
&-\frac{1}{4}\lambda^{\alpha}_+N_{+\gamma}^{\,\beta}
\Bigl[
D_{\alpha}\Fslash{\mathcal F}_{~\beta}^{\gamma}
+\frac{1}{8}\,\Fslash{\mathcal F}_{~\alpha}^{\delta}
(\gamma^{\mu}{\cal W})_{\delta}
\partial_{\mu}\Fslash{\mathcal F}_{~\beta}^{\gamma}
\Bigr]~.
\label{BRS currents on boundary}
\end{align}
Finally, we find that, to eliminate this expression, 
(\ref{DBI-alpha}), (\ref{DBI-mu}), (\ref{DBI-i}), (\ref{DBI-W}) and (\ref{DBI-PS}) should be required, as expected. 
First four equations are supersymmetric DBI equations of motion on a D$p$-brane, and last one is the pure spinor constraint.


\begin{thebibliography}{99}




\bibitem{Bosonic BI}
  E.~S.~Fradkin and A.~A.~Tseytlin,
  ``Nonlinear Electrodynamics from Quantized Strings,''
  Phys.\ Lett.\ B {\bf163} (1985) 123;\\
  A.~Abouelsaood, C.~G.~Callan Jr., C.~R.~Nappi and S.~A.~Yost,
  ``Open Strings in Background  Fields,''
  Nucl.\ Phys.\ B {\bf280} (1987) 599;\\
  R. G. Leigh, 
  ``Dirac-Born-Infeld Action from Dirichlet Sigma Model,'' 
  Mod.\ Phys.\ Lett. A {\bf4} (1989) 2767.

\bibitem{RNS BI}
  E.~Bergshoeff, E.~Sezgin, C.~N.~Pope and P.~K.~Townsend,
  ``The Born-Infeld Action from Conformal Invariance of the Open Superstring,''
  Phys.\ Lett.\ B {\bf188} (1987) 70;\\
  O.~D.~Andreev and A.~A.~Tseytlin,
  ``Partition Function Representation for the Open Superstring Effective Action: Cancellation of Mobius Infinities and Derivative Corrections to Born-Infeld Lagrangian,''
  Nucl.\ Phys.\ B {\bf311} (1988) 205;\\
  C.~G.~Callan Jr., C.~Lovelace, C.~R.~Nappi and S.~A.~Yost,
  ``Loop Corrections to Superstring Equations of Motion,''
  Nucl.\ Phys.\ B {\bf308} (1988) 221.

 \bibitem{brane kappa WZ}
  M.~Aganagic, C.~Popescu and J.~H.~Schwarz,
  ``D-Brane Actions with Local Kappa Symmetry,''
  Phys.\ Lett.\ B {\bf393} (1997) 311
  \href{http://arxiv.org/abs/hep-th/9610249}{[hep-th/9610249]};\\
  M.~Cederwall, A.~von Gussich, B.~E.~W.~Nilsson and A.~Westerberg,
  ``The Dirichlet Super Three-brane in Ten-dimensional Type IIB Supergravity,''
  Nucl.\ Phys.\ B {\bf 490} (1997) 163
  \href{http://arxiv.org/abs/hep-th/9610148}{[hep-th/9610148]};\\
  M.~Cederwall, A.~von Gussich, B.~E.~W.~Nilsson, P.~Sundell and A.~Westerberg,
  ``The Dirichlet Super $p$-branes in Ten-dimensional Type IIA and IIB Supergravity,''
  Nucl.\ Phys.\ B {\bf 490} (1997) 179
  \href{http://arxiv.org/abs/hep-th/9611159}{[hep-th/9611159]};\\  
  E.~Bergshoeff and P.~K.~Townsend,
``Super D-branes,''
  Nucl.\ Phys.\ B {\bf 490} (1997) 145
  \href{http://arxiv.org/abs/hep-th/9611173}{[hep-th/9611173]}.

 \bibitem{Cayley image}
  V.~Akulov, I.~A.~Bandos, W.~Kummer and V.~Zima,
  ``$D=10$ Dirichlet Super-9-brane,''
  Nucl.\ Phys.\ B {\bf527} (1998) 61
  \href{http://arxiv.org/abs/hep-th/9802032}{[hep-th/9802032]}.

 \bibitem{Origin embed}
  P.~S.~Howe and E.~Sezgin, 
  ``Superbranes,'' 
  Phys.\ Lett.\ B {\bf390} (1997) 133 
  \href{http://arxiv.org/abs/hep-th/9607227}{[hep-th/9607227]};\\
  P.~S.~Howe, O.~Raetzel and E.~Sezgin, 
  ``On brane actions and superembeddings,'' 
  JHEP {\bf9808} (1998) 011 
  \href{http://arxiv.org/abs/hep-th/9804051}{[hep-th/9804051]};\\
  D.~P.~Sorokin, 
  ``Superbranes and Superembeddings,'' 
  Phys.\ Rept.\ {\bf329} (2000) 1
  \href{http://arxiv.org/abs/hep-th/9906142}{[hep-th/9906142]}.

\bibitem{GS open background}
  C.~S.~Chu, P.~S.~Howe and E.~Sezgin,
  ``Strings and D-branes with Boundaries,''
  Phys.\ Lett.\ B {\bf 428} (1998) 59
  \href{http://arxiv.org/abs/hep-th/9801202}{[hep-th/9801202]}.

\bibitem{boundary fermion}
  P.~S.~Howe, U.~Lindstr\"{o}m and L.~Wulff, 
  ``Superstrings with boundary fermions,'' 
  JHEP {\bf0508} (2005) 041 
  \href{http://arxiv.org/abs/hep-th/0505067}{[hep-th/0505067]};\\
  P.~S.~Howe, U.~Lindstr\"{o}m and L.~Wulff, 
  ``On the covariance of the Dirac-Born-Infeld-Myers action,''
  JHEP {\bf0702} (2007) 070 
  \href{http://arxiv.org/abs/hep-th/0607156}{[hep-th/0607156]};\\
  P.~S.~Howe, U.~Lindstr\"{o}m and L.~Wulff, 
  ``Kappa-symmetry for coincident D-branes,'' 
  JHEP {\bf0709} (2007) 010 
  \href{https://arxiv.org/abs/0706.2494}{[arXiv:0706.2494 [hep-th]]}.




\bibitem{NADBI BRS 87}
  E.~Bergshoeff, M.~Rakowski and E.~Sezgin, 
  ``Higher Derivative Super-Yang-Mills Theories,'' 
  Phys.\ Lett.\ B {\bf185} (1987) 371.

\bibitem{NADBI T 97}
  A.~A.~Tseytlin, 
  ``On Non-abelian Generalisation of the Born-Infeld Action in String Theory,'' 
  Nucl.\ Phys.\ B {\bf501} (1997) 41 
  \href{http://arxiv.org/abs/hep-th/9701125}{[hep-th/9701125]}.

\bibitem{NADBI TR 00}
  W.~Taylor and M.~Van Raamsdonk,
  ``Multiple D$p$-branes in Weak Background Fields,''
  Nucl.\ Phys.\ B {\bf 573} (2000) 703
  \href{http://arxiv.org/abs/hep-th/9910052}{[hep-th/9910052]}.

\bibitem{NADBI K 00}
  S.~V.~Ketov, 
  ``$N=1$ and $N=2$ Supersymmetric Non-abelian Born-Infeld Actions from Superspace,'' 
  Phys.\ Lett.\ B {\bf491} (2000) 207 
  \href{http://arxiv.org/abs/hep-th/0005265}{[hep-th/0005265]}.

\bibitem{NADBI STT 01}
  A.~Sevrin, J.~Troost and W.~Troost, 
  ``The Non-Abelian Born-Infeld Action at Order $F^6$,''
  Nucl.\ Phys.\ B {\bf603} (2001) 389 
  \href{http://arxiv.org/abs/hep-th/0101192}{[hep-th/0101192]}.

\bibitem{NADBI BRS 01}
  E.~A.~Bergshoeff, M.~de~Roo and A.~Sevrin, 
  ``Non-Abelian Born-Infeld and kappa-symmetry,'' 
  J.\ Math.\ Phys. {\bf42} (2001) 2872
  \href{http://arxiv.org/abs/hep-th/0011018}{[hep-th/0011018]};\\
  E.~A. Bergshoeff, A.~Bilal, M.~de~Roo and A. Sevrin, 
  ``Supersymmetric non-abelian Born-Infeld revisited,'' 
  JHEP {\bf0107} (2001) 029 
  \href{http://arxiv.org/abs/hep-th/0105274}{[hep-th/0105274]}.

\bibitem{NADBI RSTZ 01}
  A.~Refolli, A.~Santambrogio, N.~Terzi and D. Zanon, 
  ``$F^5$ Contributions to the Non-abelian Born-Infeld Action from a Supersymmetric Yang-Mills Five-point Function,''
  Nucl.\ Phys.\ B {\bf613} (2001) 64 
  \href{http://arxiv.org/abs/hep-th/0105277}{[hep-th/0105277]}.

\bibitem{NADBI S 01}
  D.~P.~Sorokin,
 ``Coincident (super) D$p$-branes of codimension one,''
  JHEP {\bf 0108} (2001) 022
  \href{http://arxiv.org/abs/hep-th/0106212}{[hep-th/0106212]}.

\bibitem{NADBI KS 01}
  P.~Koerber and A.~Sevrin, 
  ``The non-Abelian Born-Infeld action through order $\alpha'^3$,''
  JHEP {\bf0110} (2001) 003 
  \href{http://arxiv.org/abs/hep-th/0108169}{[hep-th/0108169]};\\
  P.~Koerber and A.~Sevrin, 
  ``The non-abelian D-brane effective action through order $\alpha'^4$,''
   JHEP {\bf0210} (2002) 046 
  \href{http://arxiv.org/abs/hep-th/0208044}{[hep-th/0208044]}.

\bibitem{NADBI CRE 02}
  A.~Collinucci, M.~de~Roo and M.~G.~C.~Eenink, 
  ``Supersymmetric Yang-Mills theory at order $\alpha'^3$,''
   JHEP {\bf0206} (2002) 024
  \href{http://arxiv.org/abs/hep-th/0205150}{[hep-th/0205150]}.

\bibitem{NADBI MBM 02}
  R.~Medina, F.~T.~Brandt and F.~R.~Machado, 
  ``The open superstring 5-point amplitude revisited,''
  JHEP {\bf0207} (2002) 071
  \href{http://arxiv.org/abs/hep-th/0208121}{[hep-th/0208121]}.

\bibitem{NADBI DHL 02}
  J.~M.~Drummond, P.~S.~Howe and U.~Lindstrom,
  ``Kappa symmetric non-Abelian Born-Infeld actions in three-dimensions,''
  Class.\ Quant.\ Grav.\  {\bf 19} (2002) 6477
  \href{http://arxiv.org/abs/hep-th/0206148}{[hep-th/0206148]}.

\bibitem{NADBI DHHK 03}
  J.~M.~Drummond, P.~J.~Heslop, P.~S.~Howe and S.~F. Kerstan, 
  ``Integral invariants in $N=4$ SYM and the effective action for coincident D-branes,'' 
  JHEP {\bf0308} (2003) 016 
  \href{http://arxiv.org/abs/hep-th/0305202}{[hep-th/0305202]}.

\bibitem{NADBI CM 03}
  O.~Chandia and R.~Medina, 
  ``4-point effective actions in open and closed superstring theory,'' 
  JHEP {\bf0311} (2003) 003 
  \href{http://arxiv.org/abs/hep-th/0310015}{[hep-th/0310015]}.

\bibitem{NADBI BM 05}
  L.~A.~Barreiro and R.~Medina, 
  ``5-field terms in the open superstring effective action,''
  JHEP {\bf0503} (2005) 055 
  \href{http://arxiv.org/abs/hep-th/0503182}{[hep-th/0503182]}.





\bibitem{PS}
  N.~Berkovits,
  ``Super-Poincar\'e covariant quantization of the superstring,''
  JHEP {\bf 0004} (2000) 018
  \href{http://arxiv.org/abs/hep-th/0001035}{[hep-th/0001035]}.

\bibitem{PS open background}
  N.~Berkovits and V.~Pershin,
  ``Supersymmetric Born-Infeld from the pure spinor formalism of the open superstring,''
  JHEP {\bf 0301} (2003) 023
  \href{http://arxiv.org/abs/hep-th/0205154}{[hep-th/0205154]}.

\bibitem{Ker D9 DBI}
  S.~F.~Kerstan, 
  ``{Supersymmetric Born-Infeld from the D9-brane},'' 
  Class.\ Quant.\ Grav. {\bf 19} (2002) 4525 
  \href{http://arxiv.org/abs/hep-th/0204225}{[hep-th/0204225]}.


\bibitem{Wyl D brane}
  R.~Schiappa and N.~Wyllard, 
  ``D-brane boundary state in the pure spinor superstring,'' 
  JHEP {\bf 0507} (2005) 070 
  \href{http://arxiv.org/abs/hep-th/0503123}{[hep-th/0503123]}.

\bibitem{Gra D-brane}
  L.~Anguelova and P.~A.~Grassi,
  ``Super D-branes from BRST symmetry,''
  JHEP {\bf 0311} (2003) 010
  \href{http://arxiv.org/abs/hep-th/0307260}{[hep-th/0307260]}.

\bibitem{Muk D brane}
  P.~Mukhopadhyay, 
  ``On D-brane boundary state analysis in pure spinor formalism,''
  JHEP {\bf 0603} (2006) 066 
  \href{http://arxiv.org/abs/hep-th/0505157}{[hep-th/0505157]}.
  

 \bibitem{HS AdS}
  S.~Hanazawa and M.~Sakaguchi,
  ``D-branes from Pure Spinor Superstring in \adss{5}{5} Background,''
  Nucl.\ Phys.\ B {\bf 914} (2017) 234
  \href{https://arxiv.org/abs/1609.05457}{[arXiv:1609.05457 [hep-th]]}.
     
  \bibitem{HS mem}
  S.~Hanazawa and M.~Sakaguchi,
  ``Non-commutative M-branes from Open Pure Spinor Supermembrane,''
  Nucl.\ Phys.\ B {\bf 927} (2018) 566
  \href{https://arxiv.org/abs/1709.03711}{[arXiv:1709.03711 [hep-th]]}.

\bibitem{OT} 
  I.~Oda and M.~Tonin,
  ``On the Berkovits Covariant Quantization of GS Superstring,''
  Phys.\ Lett.\ B {\bf 520} (2001) 398
  \href{http://arxiv.org/abs/hep-th/0109051}{[hep-th/0109051]}.



\bibitem{ICTP}
  N.~Berkovits,
  ``ICTP lectures on covariant quantization of the superstring,''
  \href{http://arxiv.org/abs/hep-th/0209059}{[hep-th/0209059]}.



  \bibitem{HS4}
  S.~Hanazawa and M.~Sakaguchi,
work in progress.




\bibitem{spinorial cohomology}
  M.~Cederwall, B.~E.~W.~Nilsson and D.~Tsimpis, 
  ``The structure of maximally supersymmetric Yang-Mills theory: Constraining higher-order corrections,'' 
  JHEP {\bf0106} (2001) 034 
  \href{http://arxiv.org/abs/hep-th/0102009}{[hep-th/0102009]};\\
  M.~Cederwall, B.~E.~W.~Nilsson and D.~Tsimpis, 
  ``$D=10$ superYang-Mills at $O(\alpha'^2)$,'' 
  JHEP {\bf0107} (2001) 042 
  \href{http://arxiv.org/abs/hep-th/0104236}{[hep-th/0104236]};\\
  M.~Cederwall, B.~E.~W.~Nilsson and D.~Tsimpis, 
  ``Spinorial cohomology and maximally supersymmetric theories,'' 
  JHEP {\bf0202} (2002) 009 
  \href{http://arxiv.org/abs/hep-th/0110069}{[hep-th/0110069]}.

\bibitem{N=10 SYM const}
  B.~E.~W.~Nilsson, 
  ``Off-shell fields for ten-dimensional supersymmetric Yang-Mills theory,'' 
  Gotenburg preprint 81-6 (Feb. 1981), unpublished;\\
  B.~E.~W.~Nilsson,
  ``Pure spinors as auxiliary fields in the ten-dimensional supersymmetric Yang-Mills theory,'' 
  Class. Quant. Grav. {\bf3} (1986) L41.

\bibitem{How line}
  P.~S.~Howe, 
  ``Pure Spinor Lines in Superspace and Ten-dimensional Supersymmetric Theories,'' 
  Phys.\ Lett.\ B {\bf258} (1991) 141.

\bibitem{How function}
  P.~S.~Howe, 
  ``Pure Spinors, Function Superspace and Supergravity Theories in Ten and Eleven Dimensions,'' 
  Phys.\ Lett.\ B {\bf273} (1991) 90. 


\bibitem{HLW SC}
  P.~S.~Howe, U.~Lindstr\"{o}m and L.~Wulff, 
  ``$D=10$ supersymmetric Yang-Mills theory at $\alpha'^4$,'' 
  JHEP {\bf1007} (2010) 028 
  \href{https://arxiv.org/abs/1004.3466}{[arXiv:1004.3466 [hep-th]]};\\
  G.~Bossard, P.~S.~Howe, U.~Lindstr\"{o}m, K.~S.~Stelle and L.~Wulff,
  ``Integral invariants in maximally supersymmetric Yang-Mills theories,''
  JHEP {\bf1105} (2011) 021 
  \href{https://arxiv.org/abs/1012.3142}{[arXiv:1012.3142 [hep-th]]}.



\bibitem{Ced PS field}
  M.~Cederwall and A.~Karlsson, 
  ``Pure spinor superfields and Born-Infeld theory,'' 
  JHEP {\bf1111} (2011) 134
  \href{https://arxiv.org/abs/1109.0809}{[arXiv:1109.0809 [hep-th]]}.

\bibitem{Ced overview}
  M.~Cederwall, 
  ``Pure spinor superfields — an overview,'' 
  Springer Proc. Phys. {\bf153} (2014) 61
  \href{https://arxiv.org/abs/1307.1762}{[arXiv:1307.1762 [hep-th]]}.

\bibitem{NMPS}
  N.~Berkovits,
  ``Pure spinor formalism as an $N=2$ topological string,''
  JHEP {\bf0510} (2005) 089
  \href{http://arxiv.org/abs/hep-th/0509120}{[hep-th/0509120]};\\
  N.~Berkovits and N.~Nekrasov,
  ``Multiloop superstring amplitudes from non-minimal pure spinor formalism,''
  JHEP {\bf0612} (2006) 029
  \href{http://arxiv.org/abs/hep-th/0609012}{[hep-th/0609012]}.

\bibitem{MSYM2}
  C.~M.~Chang, Y.~H.~Lin, Y.~Wang and X.~Yin, 
  ``Deformations with maximal supersymmetries part 2: off-shell formulation,''
   JHEP {\bf1604} (2016) 171
   \href{https://arxiv.org/abs/1403.0709}{[arXiv:1403.0709 [hep-th]]}.

\bibitem{Ber Cederwall action}
  N.~Berkovits and M.~Guillen, 
  ``Equations of motion from Cederwall's pure spinor superspace actions,''
  JHEP {\bf1808} (2018) 033
  \href{https://arxiv.org/abs/1804.06979}{[arXiv:1804.06979 [hep-th]]}.




\bibitem{PS closed background}
  N.~Berkovits and P.~S.~Howe,
  ``Ten-dimensional Supergravity Constraints from the Pure Spinor Formalism for the Superstring,''
  Nucl.\ Phys.\ B {\bf 635} (2002) 75
  \href{http://arxiv.org/abs/hep-th/0112160}{[hep-th/0112160]}.

\bibitem{GS closed background}
  M.~Grisaru, P.~S.~Howe, L.~Mezincescu, B.~E.~W.~Nilsson and P.~K.~Townsend,
  ``$N=2$ Superstrings in a Supergravity Background,''
  Phys.\ Lett.\ B {\bf 162} (1985) 116.

  

\bibitem{TseWul SSE}
  A.~A.~Tseytlin and L.~Wulff, 
  ``Kappa-symmetry of superstring sigma model and generalized 10d supergravity equations,'' 
  JHEP {\bf1606} (2016) 174 
  \href{https://arxiv.org/abs/1605.04884}{[arXiv:1605.04884 [hep-th]]}.

\bibitem{GSEDFT}
  Y.~Sakatani, S.~Uehara and K.~Yoshida, 
  ``Generalized gravity from modified DFT,'' 
  JHEP {\bf1704} (2017) 123 
  \href{https://arxiv.org/abs/1611.05856}{[arXiv:1611.05856 [hep-th]]};\\
  A.~Baguet, M.~Magro and H.~Samtleben, 
  ``Generalized IIB supergravity from exceptional field theory,'' 
  JHEP {\bf1703} (2017) 100 
  \href{https://arxiv.org/abs/1612.07210}{[arXiv:1612.07210 [hep-th]]};\\
  J.~Sakamoto, Y.~Sakatani and K.~Yoshida, 
  ``Weyl invariance for generalized supergravity backgrounds from the doubled formalism,'' 
  PTEP {\bf2017} (2017) no. 5, 053B07
  \href{https://arxiv.org/abs/1703.09213}{[arXiv:1703.09213 [hep-th]]}.

\bibitem{generalized SUGRA}
  G.~Arutyunov, S.~Frolov, B.~Hoare, R.~Roiban and A.~A.~Tseytlin, 
  ``Scale Invariance of the $\eta$-deformed $AdS_5 \times S^5$ Superstring, T-duality and Modified Type II Equations,'' 
  Nucl. Phys. B {\bf903} (2016) 262 
  \href{https://arxiv.org/abs/1511.05795}{[arXiv:1511.05795 [hep-th]]}.


 
\bibitem{Wit Sie SYM}
  W.~Siegel, ``{Superfields in Higher Dimensional Spacetime},'' Phys. Lett. B {\bf 80} (1979) 220;\\
  E.~Witten, ``{Twistor-like Transform in Ten Dimensions},'' Nucl. Phys. B {\bf 266} (1986) 245.


\bibitem{N=4 SYM const}
  M.~Sohnius,
  ``{Bianchi Identities for Supersymmetric  Theories},'' 
  Nucl. Phys. B {\bf136} (1978) 461.

\bibitem{theta expansion}
  J.~P.~Harnad and S.~Shnider, 
  ``{Constraints and Field Equations for Ten-Dimensional Super Yang-Mills Theory},'' 
  Commun. Math. Phys. {\bf106} (1986) 183;\\
  H.~Ooguri, J.~Rahmfeld, H.~Robins and J.~Tannenhauser, 
  ``{Holography in superspace},'' 
  JHEP {\bf0007}  (2000) 045 
  \href{http://arxiv.org/abs/hep-th/0007104}{[hep-th/0007104]};\\
  P.~A.~Grassi and L.~Tamassia, 
  ``{Vertex operators for closed superstrings},'' 
  JHEP {\bf0407} (2004) 071 
  \href{http://arxiv.org/abs/hep-th/0405072}{[hep-th/0405072]};\\
  G.~Policastro and D.~Tsimpis, 
  ``{$R^4$, purified},'' 
  Class. Quant. Grav. {\bf23} (2006) 4753
  \href{http://arxiv.org/abs/hep-th/0603165}{[hep-th/0603165]}.





\bibitem{Bak b}
  I.~Bakhmatov and N.~Berkovits,
  ``Pure spinor $b$-ghost in a super-Maxwell background,''
  JHEP {\bf1311} (2013) 214
  \href{https://arxiv.org/abs/1310.3379}{[arXiv:1310.3379 [hep-th]]}.




 
 
 
\end{thebibliography}
\end{document}